\documentclass[12pt,preprint]{aastex}

\newcommand{\gcc}{\mathrm{g~cm^{-3} }}

\usepackage{epsf,color,dcolumn}
\newcolumntype{d}{D{.}{.}{-1}}

\epsfverbosetrue

\begin{document} 

\title{Three-dimensional Numerical Simulations of Rayleigh-Taylor
Unstable Flames in Type Ia Supernovae}

\shorttitle{3d Simulations of RT Unstable SNe Ia flames}
\shortauthors{Zingale et al.}

\author{M.    Zingale\altaffilmark{1},
        S.~E. Woosley\altaffilmark{1},
        C.~A. Rendleman\altaffilmark{2},
        M.~S. Day\altaffilmark{2},
        J.~B. Bell\altaffilmark{2}}

\altaffiltext{1}{Dept. of Astronomy \& Astrophysics,
                 The University of California, Santa Cruz,
                 Santa Cruz, CA 95064}

\altaffiltext{2}{Center for Computational Science and Engineering,
                 Lawrence Berkeley National Laboratory,
                 Berkeley, CA 94720}

\begin{abstract}

Flame instabilities play a dominant role in accelerating the burning
front to a large fraction of the speed of sound in a Type~Ia
supernova.  We present a three-dimensional numerical simulation of a
Rayleigh-Taylor unstable carbon flame, following its evolution through
the transition to turbulence.  A low Mach number hydrodynamics method
is used, freeing us from the harsh timestep restrictions imposed by
sound waves.  We fully resolve the thermal structure of the flame and
its reaction zone, eliminating the need for a flame model.  A single
density is considered, $1.5\times 10^7~\gcc$, and half carbon\slash
half oxygen fuel---conditions under which the flame propagated in the
flamelet regime in our related two-dimensional study.  We compare to a
corresponding two-dimensional simulation, and show that while
fire-polishing keeps the small features suppressed in two dimensions,
turbulence wrinkles the flame on far smaller scales in the
three-dimensional case, suggesting that the transition to the
distributed burning regime occurs at higher densities in three
dimensions.  Detailed turbulence diagnostics are provided.  We show
that the turbulence follows a Kolmogorov spectrum and is highly
anisotropic on the large scales, with a much larger integral scale in
the direction of gravity.  Furthermore, we demonstrate that it becomes
more isotropic as it cascades down to small scales.  Based on the
turbulent statistics and the flame properties of our simulation, we
compute the Gibson scale.  We show the progress of the turbulent flame
through a classic combustion regime diagram, indicating that the flame
just enters the distributed burning regime near the end of our
simulation.
\end{abstract}


% ============================================================================
%  Introduction
% ============================================================================
\section{INTRODUCTION}
\label{sec:intro}

The importance of the Rayleigh-Taylor~(RT) instability and turbulence
in accelerating a thermonuclear flame in Type~Ia supernovae (SNe~Ia)
is well recognized
\citep{mullerarnett1982,livne1993,khokhlov1995,niemeyerhillebrandt1995b,gamezo2003,reinecke2002b}.
Nevertheless, the physics of reactive flame instabilities in SNe~Ia is
still not completely understood.  In a previous study \citep{SNrt}, we
showed the transition of the burning front from the flamelet regime to
the distributed burning regime occurred at a density of $\sim
10^7~\gcc$, through two-dimensional direct numerical simulations.
This transition had long been postulated, but these simulations were
the first to demonstrate it.  It was recently shown that large scale
models may incorporate a specialized prescription for distributed burning
within the flame model to release more energy, produce higher-velocity ejecta
\citep{roepkehillebrandt2004}, and bring them more in line with
observations.  But the arbitrary nature of these prescriptions
underscores the importance of a firm understanding
of the distributed burning regime.  In the present paper, we extend
our study to three dimensions while continuing to fully resolve the
thermal structure of the flame.  Due to the computational expense of
three-dimensional simulations, we concentrate on a single density,
$1.5\times 10^7~\gcc$.  In our two-dimensional study, this was in the
flamelet regime.  In three dimensions, turbulence is expected to play
a bigger role \citep{khokhlov1994}, and wrinkling on even smaller
scales could make the transition density to the distributed regime
higher in three dimensions than it is in two dimensions.  We note that
the transition to distributed burning occurs when portions of the
flame smaller than the flame thickness become wrinkled.  This can
occur either through turbulence or the RT~instability directly, as
demonstrated in our two-dimensional study.

Avoiding the use of a flame model, such as front-tracking
\citep{reinecke1999a} or thickened flames \citep{khokhlov1993,
khokhlov1994}, allows us to understand the small scale physics of the
flame with no prior assumptions.  There is no need to set a flame
speed---the input physics yield the proper speed automatically, with
proper coupling between the flame and the flow.  Nor is any subgrid
flame model required.  In the flamelet regime, the burning sets the
smallest physical scale.  The importance of curvature in modifying the
local burning rate \citep{flame-curvature} is implicitly accounted
for.  When turbulence becomes important, scales down to the Gibson
scale---the scale at which a turbulent eddy burns away before turning
over---need to be followed for a resolved calculation, otherwise, some
sort of turbulence subgrid model is likely required.  Resolving the
thermal structure does constrain the size of the region we can
consider to $\sim 100$~flame thicknesses, due to computational
resources.  Furthermore, to see RT~growth, our domain must be several
times wider than the small scale cutoff for reactive~RT, namely, the
fire-polishing length,
\begin{equation}
\label{eq:firepolishing}
\lambda_\mathrm{fp} = 4 \pi \frac{{S_l}^2}{g_{\mathrm{eff}}} \enskip ,
\end{equation}
\citep{timmeswoosley1992} where $g_{\mathrm{eff}}$ is the effective
gravitational acceleration and $S_l$ is the laminar flame speed.
Since the fire-polishing length grows and the flame thickness
decreases with increasing density, these constraints make the region
centered around $10^7~\gcc$ the optimal density for direct numerical
simulations.  This is illustrated in Figure~\ref{fig:regimes}, where
we plot the flame thickness and the fire-polishing lengths as a
function of density.  We see these lines cross just between $\rho =
10^7~\gcc$ and $\rho = 1.5\times 10^7~\gcc$.  The shaded region
indicates a factor of two uncertainty in the fire-polishing length.
%The data for this figure was computed with the present simulation
%code.

Some measurements of the response of the local burning rate to
curvature and strain were provided in one-dimensional spherical flame
studies \citep{flame-curvature} and in our previous study of the
Landau-Darrieus instability \citep{SNld}.  Here we can assess the
difference in the effect of curvature between two and three
dimensions, where the surface to volume ratio of the RT~unstable flame
changes significantly.  These small-scale flame studies provide
important information on flame physics in SNe~Ia that can be used to
build a subgrid model for large scale flame simulations where it is
no longer possible to resolve the flame structure.

Modeling the flame in three dimensions will also allow us to gain an
understanding of the nature of the turbulence.  The Kelvin-Helmholtz
instabilities accompanying the RT~instability in SNe~Ia drives most of
the turbulence in the star, and, as the flame wrinkles, it will
interact with the turbulence generated on larger scales.
Three-dimensional simulations of the flame can give us an idea about
the nature of this flame-generated turbulence.  Specifically, is it
isotropic? Does it obey Kolmogorov or Bolgiano-Obukhov statistics
\citep{niemeyerkerstein1997,chertkov2003}?  Turbulent subgrid models
are the foundation of some large scale SNe~Ia simulations
\citep{niemeyerhillebrandt1995b}, but these models always assume the turbulence
to be isotropic.  Any anisotropy can have a large effect on the
progression of the flame front.  If the flame burns faster radially
than laterally, as initial hot spots rise, these burning spots may not
merge, leaving behind unburned fuel.  An important question is over
what scales does the turbulence remain anisotropic?

Because of the computational demands of three dimensions, the size of
the domain that we can consider is limited, and we use large
two-dimensional simulations to complement the results of the
three-dimensional calculations.  This study will give us an idea of
how good an approximation the two-dimensional simulations are.  Only a
single density is considered here, $1.5\times 10^7~\gcc$, with half
carbon\slash half oxygen fuel.  We also compare our results with the
thickened-flame model RT~simulations performed by
\citet{khokhlov1995}.  Finally, these results can be used to
directly validate different flame models---a process that is
underway, but beyond the scope of the present paper.

In \S~2, we outline the basic formulation of our simulation code and
input physics used for the present simulations.  Results of the two-
and three-dimensional RT~simulations are presented in \S~3, and we
conclude in \S~4 with a comparison to previous three-dimensional
reactive RT~simulations.

% ============================================================================
%  Numerical Methods
% ============================================================================
\section{SIMULATION DETAILS}
\label{sec:numerics}

The numerical method is identical to that described in detail in
\cite{SNeCodePaper}, and used in our two-dimensional RT~study
\citep{SNrt}, so we only briefly summarize here.  A low Mach number
formulation of the inviscid Navier-Stokes equations is solved using
an advection\slash projection\slash reaction formalism.  In this approximation,
the pressure is decomposed into a dynamic and thermodynamic component,
the ratio of which is $O(M^2)$.  The thermodynamic component is assumed
to be spatially uniform, so that only the dynamic component appears
in the momentum equation.  As a result, acoustic waves are not supported
analytically.  The advection algorithm is a conservative
second-order accurate unsplit Godunov method, which yields better
symmetry and resistance to grid imprinting than corresponding
directionally split methods.  An elliptic constraint on the velocity
field, derived by requiring the thermodynamic pressure to be constant
along streamlines, is solved in the projection step.  The advantage of
the low Mach number formulation is that that the timesteps depend
solely on the fluid velocity and not the fluid velocity plus sound
speed.  As the flames we consider here are $M \ll 1$, this greatly
reduces the number of timesteps required, although due to the
projection step each timestep requires more computational resources.
The reaction step uses a single rate for the
$^{12}\mathrm{C}(^{12}\mathrm{C},\gamma)^{24}\mathrm{Mg}$ reaction
from \citet{caughlan-fowler:1988}.  The thermal conductivities are
those from \citet{timmes_he_flames:2000}, and the equation of state is
appropriate for electron degenerate matter, as provided in
\citet{timmes_swesty:2000}.  The code is parallelized using MPI,
distributing patches in the mesh hierarchy across processors in a
load-balanced fashion \citep{rendleman2000}.  The adaptive mesh
refinement (AMR) follows the strategy outlined in \citet{bergercolella}.

Only a single density is considered here, $1.5\times 10^7~\gcc$, with
the fuel half carbon and half oxygen, which, in our two-dimensional
study, was comfortably in the flamelet regime.  As in the
two-dimensional study, only the carbon is burned.  The laminar flame
speed, $S_l$, is $8.19\times 10^3~\mathrm{cm~s^{-1}}$ and the thermal
flame thickness, $\delta$, is $0.55$ to $1.25~\mathrm{cm}$ depending
on the definition.  The former corresponds to
\begin{equation}
\label{eq:ft1}
\delta = \frac{T_\mathrm{ash}-T_\mathrm{fuel}}{\max \{ | \nabla T| \} } \enskip ,
\end{equation}
common in terrestrial combustion, while the later is 
\begin{equation}
\label{eq:ft2}
\delta = x(T = 0.9 \max \{T\}) - x(T = 0.1 \max \{T\}) \enskip ,
\end{equation}
as used in \citep{timmeswoosley1992}.  This factor of two uncertainty
makes some definitions of the burning regime the flame is operating in
uncertain.  Unless otherwise indicated, we will use the former
definition.  The Atwood number of this flame is 0.28.

The base grid for this simulation is $32\times 32\times 64$, and
covers the entire domain.  Four additional levels of adaptive mesh
refinement are used, and track the location of the flame surface.
Each successive level is a factor of 2
finer, yielding an effective fine grid of $512\times 512\times 1024$
zones.  We note that this grid is quite large even for
non-reactive RT~simulations.
A recent study comparing several codes simulating the pure
RT~instability
\citep{alphagroup} used grids no larger than $256\times
256\times 512$, and found this to be more than adequate to observe the
full nonlinear phase of the RT~instability development.  As we will show in
\S~\ref{sec:results}, with this size grid we capture the
turbulent cascade over more than one and a half orders of magnitude.
The domain is $53.5~\mathrm{cm} \times 53.5~\mathrm{cm} \times
107.0~\mathrm{cm}$---the same size as the corresponding
two-dimensional simulation from \citet{SNrt}, yielding a grid
resolution of $0.1$~cm.  This puts just over 5~zones in the flame
thickness, which in two-dimensional studies was enough to yield
convergence of the flame speed, length, and width \citep{SNrt}.

%In three-dimensions, turbulence becomes important, and resolving the
%dissipation scale is required for a resolved simulation.  In a white
%dwarf, this scale is quite small, and as a result, we solve the
%inviscid Navier-Stokes equations.
Although momentum transport in a white dwarf is dominated by convective 
processes so that we may safely neglect physical viscosity terms, 
burning sets a lower bound on the scale length of the turbulence.
The associated length, the Gibson scale, should be resolved
on the grid.  It is difficult to predict the
Gibson scale length, since it depends on the nature of the turbulence, but
estimates suggest it to be around 1.0~cm
\citep{niemeyerwoosley1997}.
%As we demonstrate in
%\S~\ref{sec:results} we do resolve on our grid.  This means that the
Based on this estimate, the smallest turbulent eddies
that can affect our flame are resolved on our grid.  The
fire-polishing wavelength is $3.0$~cm, so we more than resolve the
smallest scale that RT will grow on.

To initialize the problem, a steady-state laminar flame at this
density is mapped onto the domain, and shifted about its zero point
by a perturbation,
\begin{equation}
\Delta = A \sum_{m=1}^{M} \sum_{n=1}^{N} \delta_{m,n} \enskip ,
\end{equation}
where
\begin{equation}
\delta_{m,n} = a_{m,n} \sin\left(\frac{2\pi m x}{L_x} +
\phi^x_{m,n}\right) \sin\left(\frac{2\pi n y}{L_y} +
\phi^y_{m,n}\right) \enskip ,
\end{equation}
with $\phi^x_{m,n}$ and $\phi^y_{m,n}$ chosen randomly, and $a_{m,n}$
chosen randomly to fall between -1 and 1.  Here, $L_x$ and $L_y$ are
the size of the domain in the $x$- and $y$-direction respectively.
The overall amplitude, $A$ is set to make the maximum perturbation
$\Delta = 1~ \mathrm{cm}$.  For the present simulation, we pick $M = N
= 10$.  At the lower boundary ($-$Z), fuel enters the domain at the laminar
flame speed.  The upper boundary ($+$Z) is outflow, and the transverse
directions are periodic.  Because of the inflow conditions, an
unperturbed flame would remain stationary in our domain.  Gravity
points in the $z$-direction, and is taken as
$10^9~\mathrm{cm~s^{-2}}$.

The run begins relatively unrefined, with the fine grid taking up less
than 10\% of the domain, but as the non-linear~RT develops, the number
of zones grows to over 160~million, with the fine grid covering almost half of
the domain.  In all, 395~coarse grid timesteps were taken, corresponding to
approximately 1.2~ms of physical evolution.  The
simulation was stopped when the RT~spikes began to approach the upper
boundary.  The 2:1 aspect ratio of our domain means that this will
occur when the mixed region becomes larger than the width of the
domain, after this point there are no more modes that can grow in the
periodic domain.  For comparison, the identical setup is run in two
dimensions.

% ============================================================================
%  Results 
% ============================================================================
\section{RESULTS}
\label{sec:results}

Figure~\ref{fig:3dfull} shows a volume rendering of the carbon mass
fraction at several instances in time.  Throughout this section, we will
refer to them by the letter key for each panel.
As the simulation begins, the initial perturbations burn away
prior to the linear-growth phase of the RT~instability.
The seeded modes grow strongly, creating a highly wrinkled flame surface.
The initial growth of the RT~instability and the smooth fire-polished
features between the characteristic RT~mushroom caps looks very
similar to the early two-dimensional evolution, as shown in
Figure~\ref{fig:2dview}.  These growing bubbles begin to interact and
merge, creating a more extended mixed region.  In
Figures~\ref{fig:3dfull}c,d, these modes begin to break
down as they become wrinkled on increasingly smaller scales.  This provides
a first indication that turbulence is operating on scales smaller
than the fire-polishing length.  At late times,
Figure~\ref{fig:3dfull}e,f, we see a very chaotic flame.  It has 
become quite thick and it is hard to discern the original modes we
seeded, as they have broken down into a more turbulent flame.  This
stage of the flame evolution has no two-dimensional analogue.
% The simulation was stopped when the flow approached the 
% top boundary of our domain.

\begin{figure*}
\begin{center}
\epsscale{1.0}
\plotone{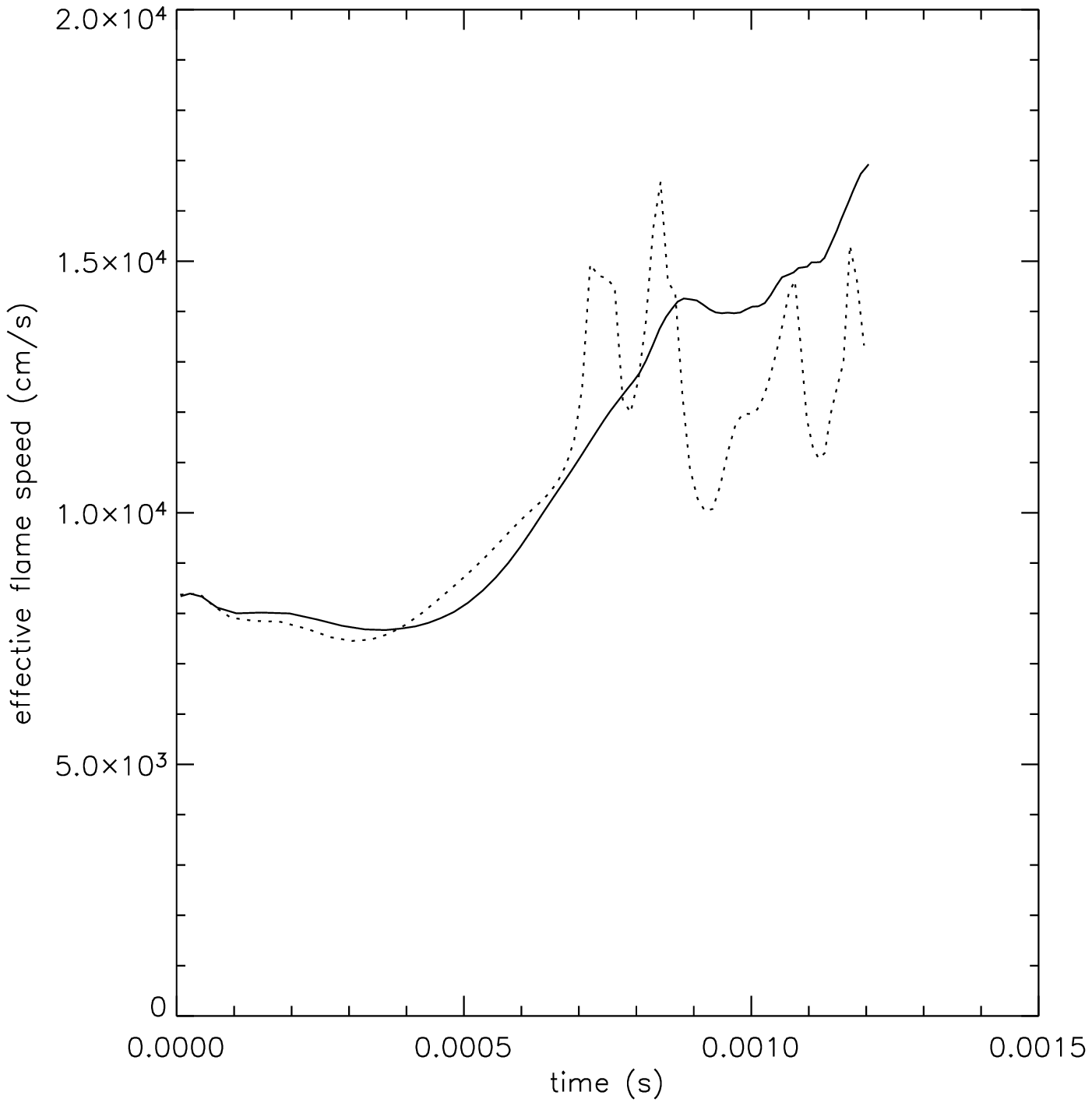}
\epsscale{1.0}
\end{center}
\caption{\label{fig:speeds} Effective flame speed for the 2d (dotted)
and 3d (solid) flame simulations.  The acceleration at late times in
the 3d simulation is due to the onset of turbulence.}
\end{figure*}

\begin{figure*}
\begin{center}
\epsscale{1.0}
\plotone{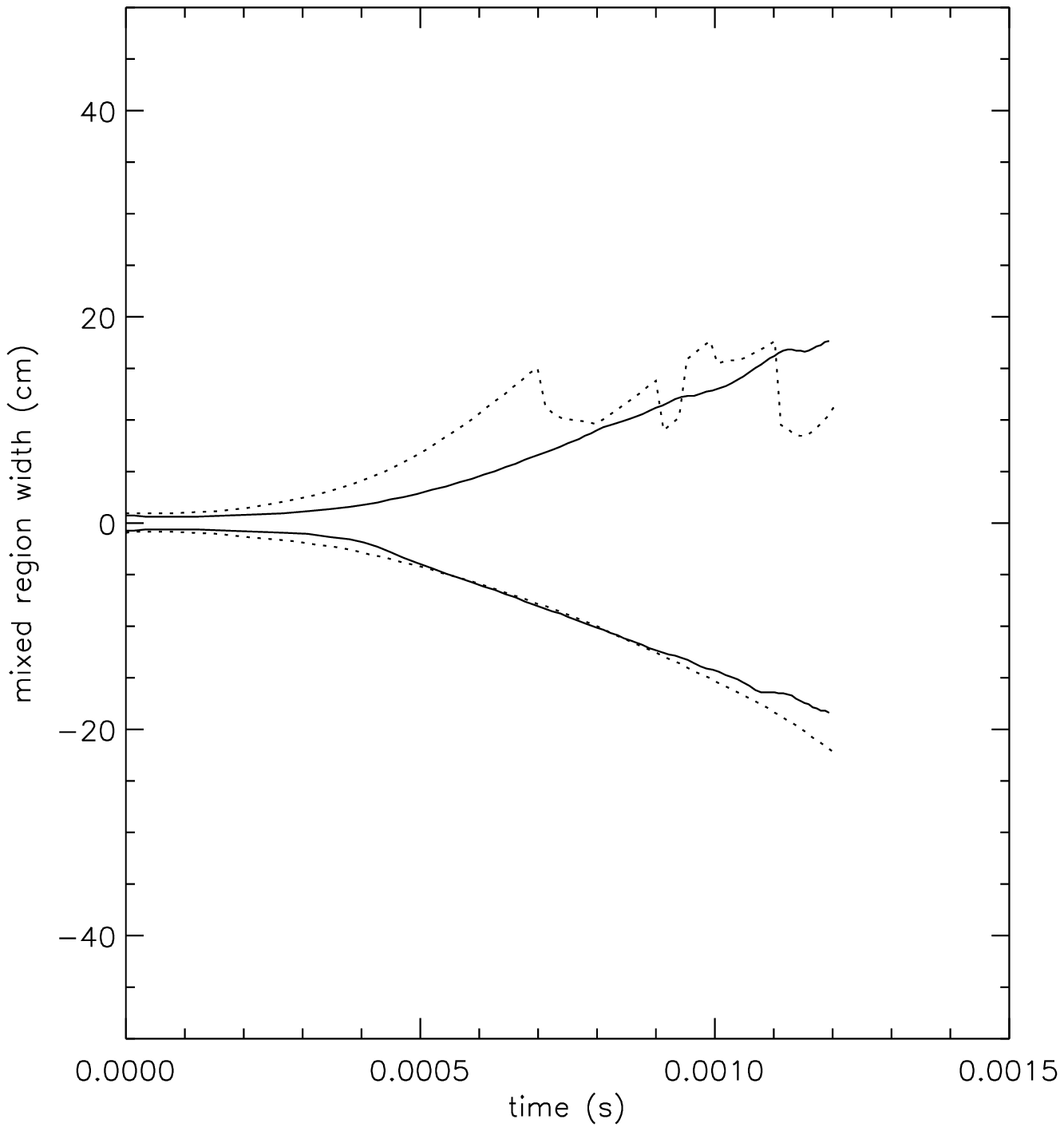}
\epsscale{1.0}
\end{center}
\caption{\label{fig:width} Extent of the mixed region for the 2d
(dotted) and 3d (solid) flame simulations.  We see that the position
of the spikes of fuel moving into the hot ash (top curves) is greatly
suppressed in three dimensions, owing to the greater surface to volume
area enhancing the burning.}
\end{figure*}

Figures~\ref{fig:speeds}--\ref{fig:width} show the flame speed,
relative change in surface area, and width of the mixed region for the
two- and three-dimensional runs.  The effective flame speed is
computed by looking at the carbon consumption rate on the domain, as
detailed in \citet{SNrt}.
The flame surface area is computed by evaluating the area (length
in two dimensions) for instantaneous isosurfaces of constant carbon mass fraction.
% MARC thinks it better not to say marching cubes.  Alternatively, you
% can find a reference...
% The surface area is computed using the
% marching cubes algorithm (marching squares in two dimensions), seeking
% a contour of constant carbon mass fraction.
The carbon mass fraction may be averaged laterally across the domain to
define a mean progress variable as a function of $z$ only.  The width
of the mixed region is then defined as the distance separating the
resulting $0.05$ and $0.45$ elevations of the mean progress.

We see a similar trend in the effective flame speed
(Figure~\ref{fig:speeds}) in two and three dimensions in the early
stages of evolution.  The curve for three dimensions is smoother,
since there are more modes growing (due to the extra dimension) and
therefore the effect of a single spike of fuel burning away is
lessened.  Both curves show an initial decline in the flame speed, as
the initial perturbations start to burn away before the RT~growth sets
in.  After about 0.0005~s, the RT~begins to set in strongly, and the
flame speed accelerates to about twice the laminar value.  However, at
late times, we see the effects of the turbulence in three dimensions,
and the flame speed accelerates again.  In the two-dimensional case, 
the flame speed saturates after about $0.7$~ms.  The
saturation is likely due to the finite-size domain, and a higher peak value
would be observed in a larger domain.

\begin{figure*}
\begin{center}
\plottwo{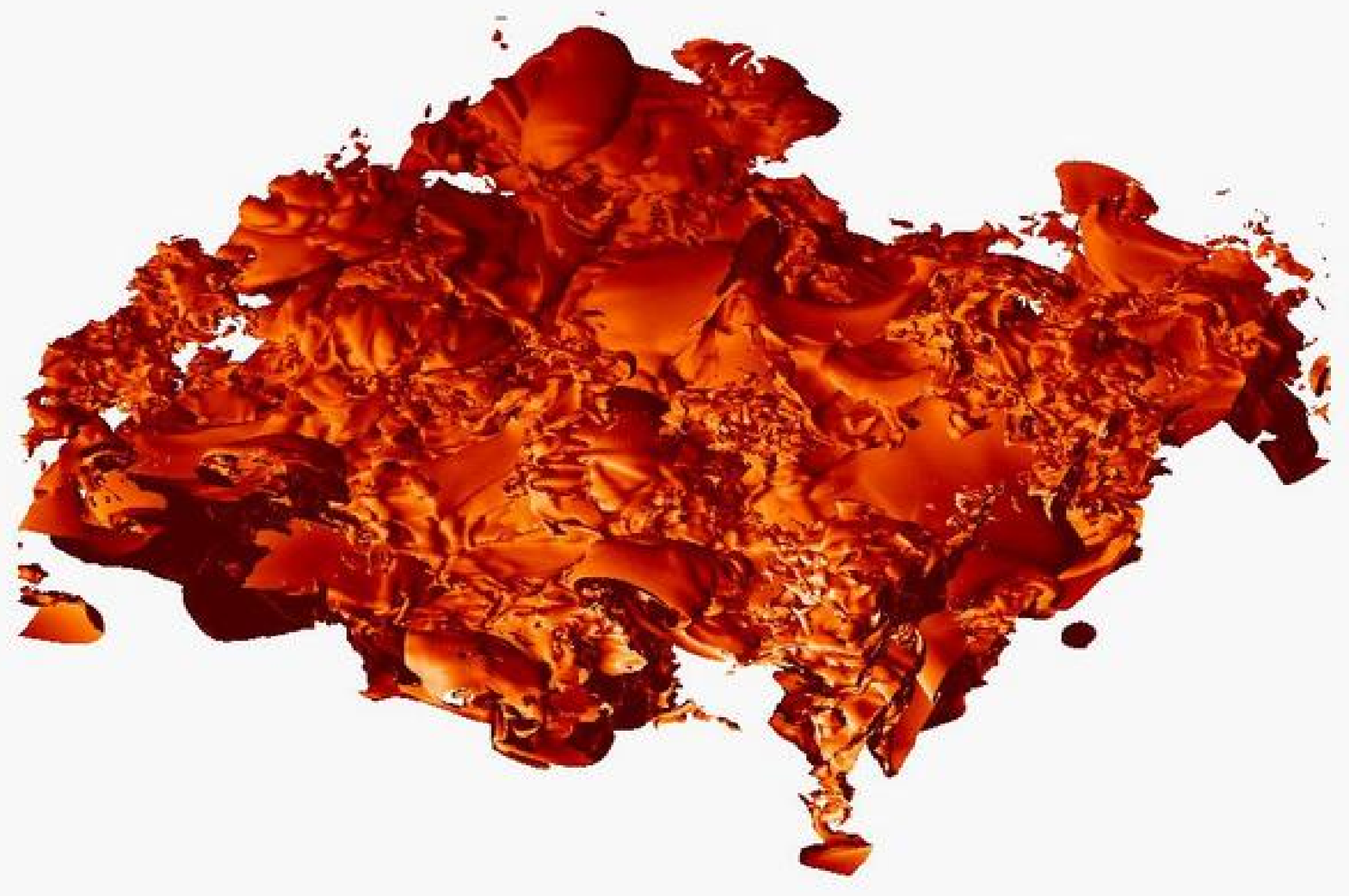}{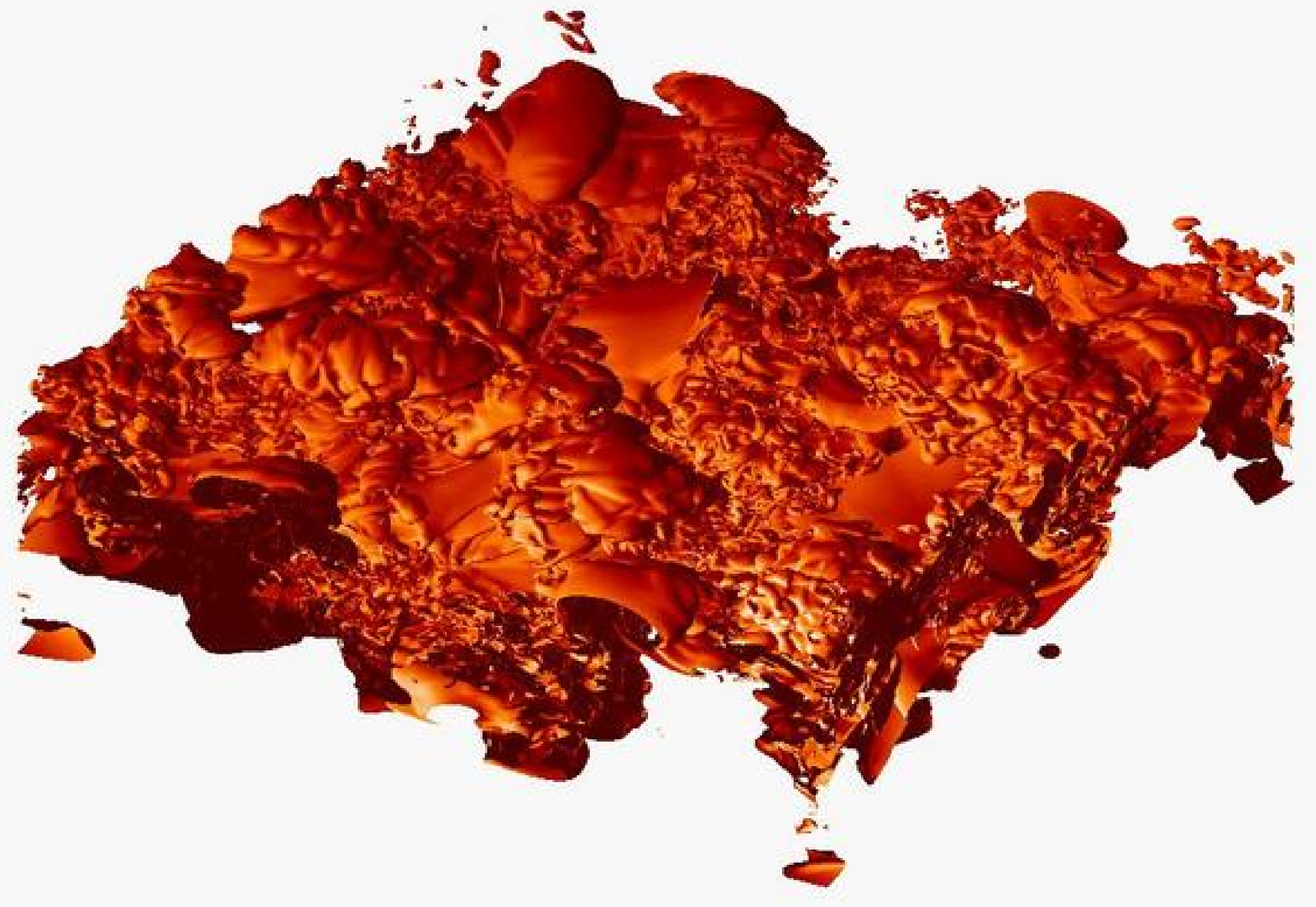}
\end{center}
\caption{\label{fig:isovaluecompare} Isosurfaces of the carbon mass
fraction of 0.1 (left) and 0.25 (right) at $10^{-3}$~s.}
\end{figure*}

Figure~\ref{fig:length} shows the surface area evolution for the two
runs.  For each case, two curves are shown, using carbon
mass fraction values of~0.25 and~0.1.  To allow for a comparison
between the two- and three-dimensional runs, we plot the
fractional difference of the area,
\begin{equation}
f = \frac{A(t) - A_0}{A_0} \enskip ,
\end{equation}
where $A_0$ is the cross-sectional area of the domain.  For both
dimensions, we see a delay in the growth of the flame surface area as
the burning eats away at the initial perturbations.  Nevertheless, the
linear growth of the RT~instability quickly sets in, and we see the
flame surface area grow faster in the three-dimensional case than the
corresponding two-dimensional simulation after 0.0004~s.  The surface
area is somewhat sensitive to the value of the carbon mass fraction
being used, differing by about 25\% at late times in the
three-dimensional case.  Isosurfaces of the two mass fraction values
at a representative time of $t=0.1$~ms are shown in
Figure~\ref{fig:isovaluecompare}.  The $0.25$-surface exhibits a
higher level of fine-scale structure, resulting in a larger surface area.
However, both surfaces exhibit similar trends.  By the end of the simulation,
the three-dimensional case shows a factor of $12$-$15$ enhancement in
flame area.

The width of the mixed region shows the effect of this greater surface
to volume ratio in three dimensions clearly.  Each simulation has two
curves, the interface of the fuel and the mixed region (the top curve
for each simulation) and the interface of the ash and the mixed
region.  The fuel\slash mixed interface appears choppy, because this is
where the spikes of fuel are falling into the hot ash and rapidly
burning away.  In three dimensions, these spikes of fuel burn away
much more rapidly, and, as Figure~\ref{fig:width} shows, this
interface grows much slower in three dimensions than two dimensions.
This is to be contrasted with non-reactive RT~simulations
(see for example \citealt{young2001,flash-validation,alphagroup})
that show the three-dimensional case growing faster than the
two-dimensional case.  When reactions are added, the increase in the
surface to volume ratio of the spikes of fuel makes the
three-dimensional growth slower than the corresponding two-dimensional
case.  At late times, once turbulence has set in, we see a departure
of the bubbles of ash pushing into the cool fuel from the smooth curve
characteristic of the RT~instability observed in two dimensions.

\begin{figure*}
\begin{center}
\plotone{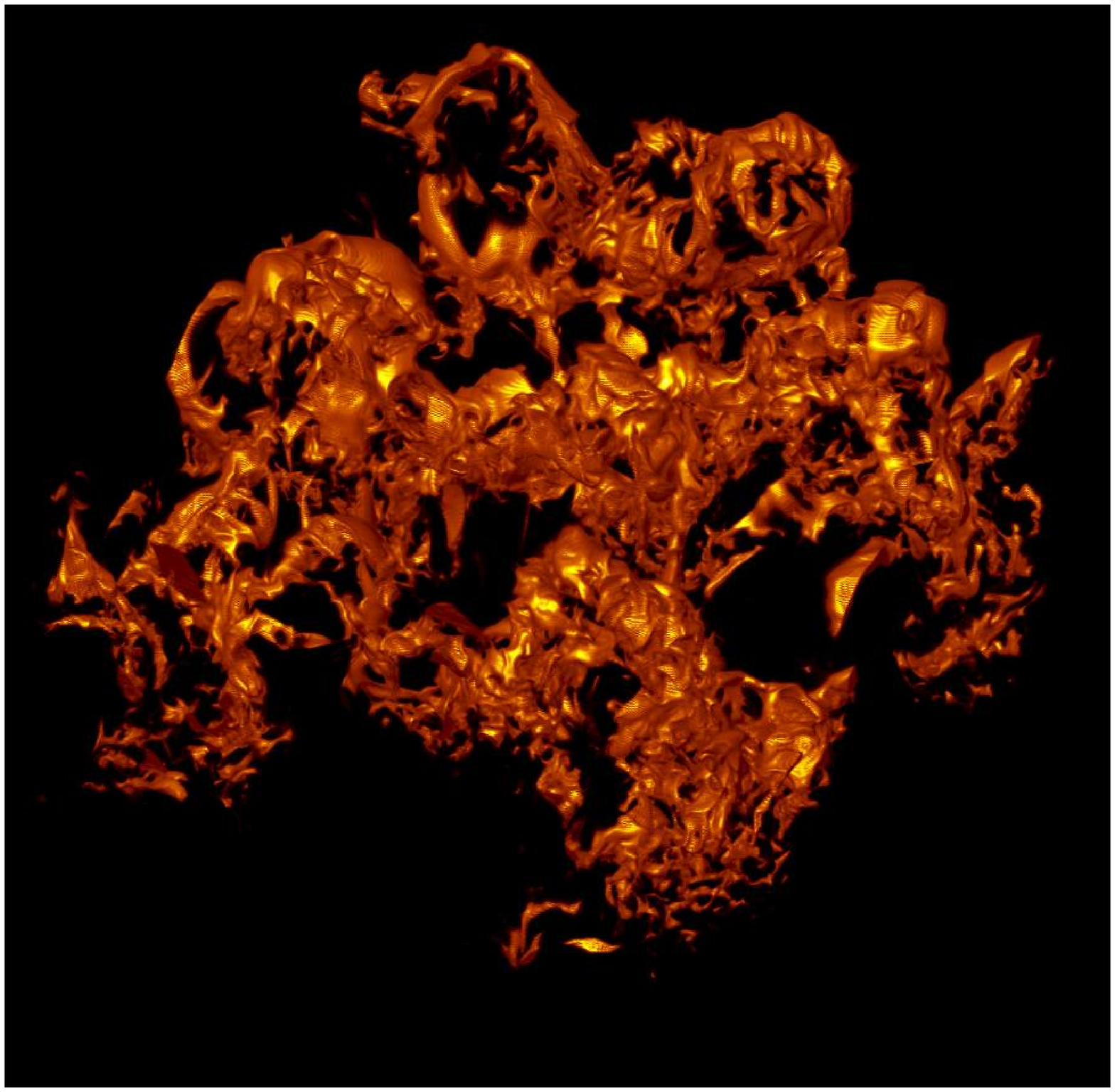}
\end{center}
\caption{\label{fig:enuc} Volume rendering of the carbon destruction
rate at $1.16\times 10^{-3}$~s, showing the reacting surface is not
uniform.}
\end{figure*}

\begin{figure*}
\begin{center}
\plotone{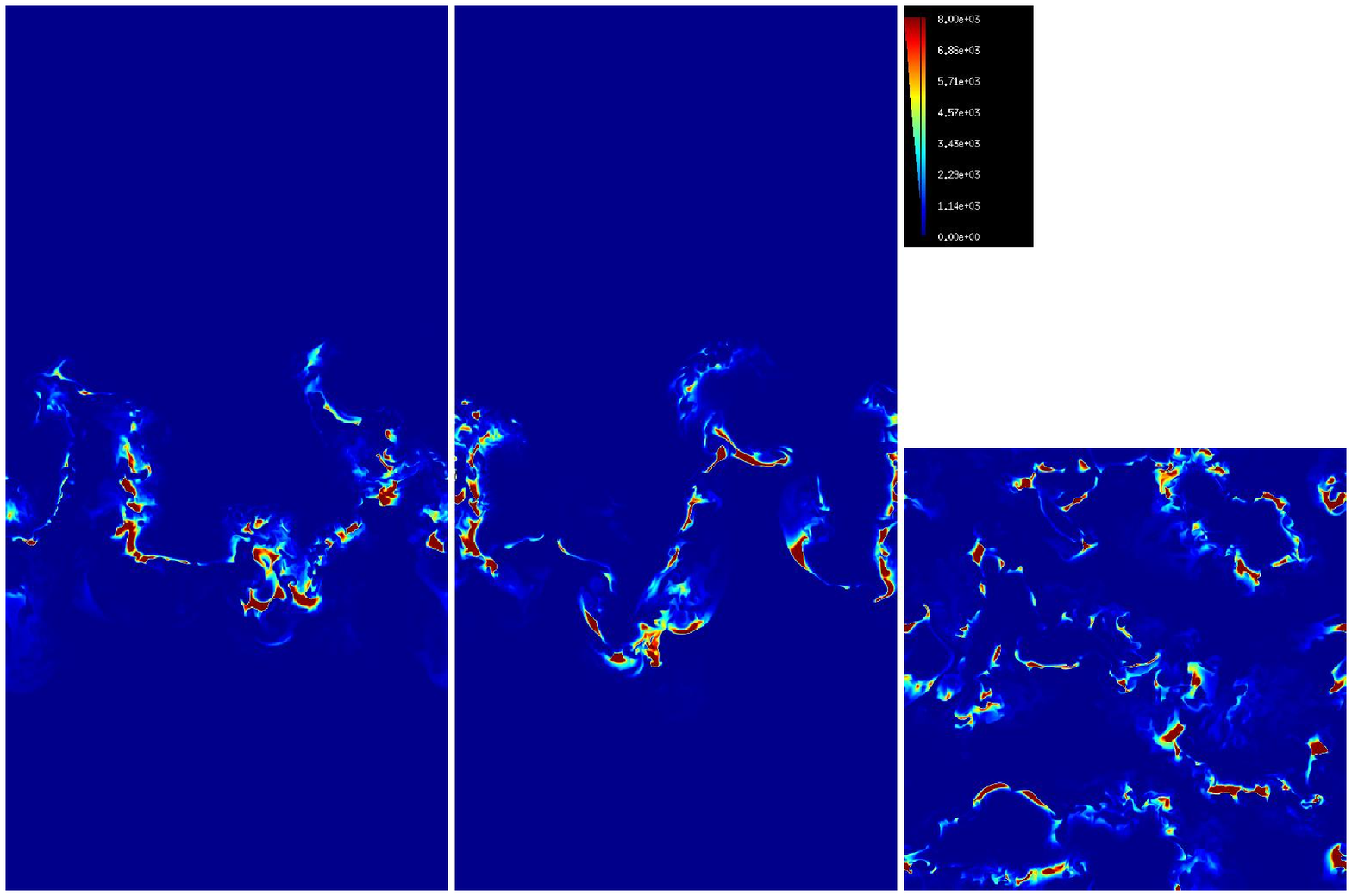}
\end{center}
\caption{\label{fig:enuc_slices} Slices of the absolute value of the
carbon destruction rate at $1.16\times 10^{-3}$~s, for the
three-dimensional calculation.  Shown at $x-z$ (left), $y-z$ (center),
and $x-y$ (right).  These provide an alternate view to the volume
rendering in Figure~\ref{fig:enuc}.  In this plot, the limits of the
colormap are chosen to match the laminar burning rates.}
\end{figure*}

\begin{figure*}
\begin{center}
\epsscale{0.8}
\plotone{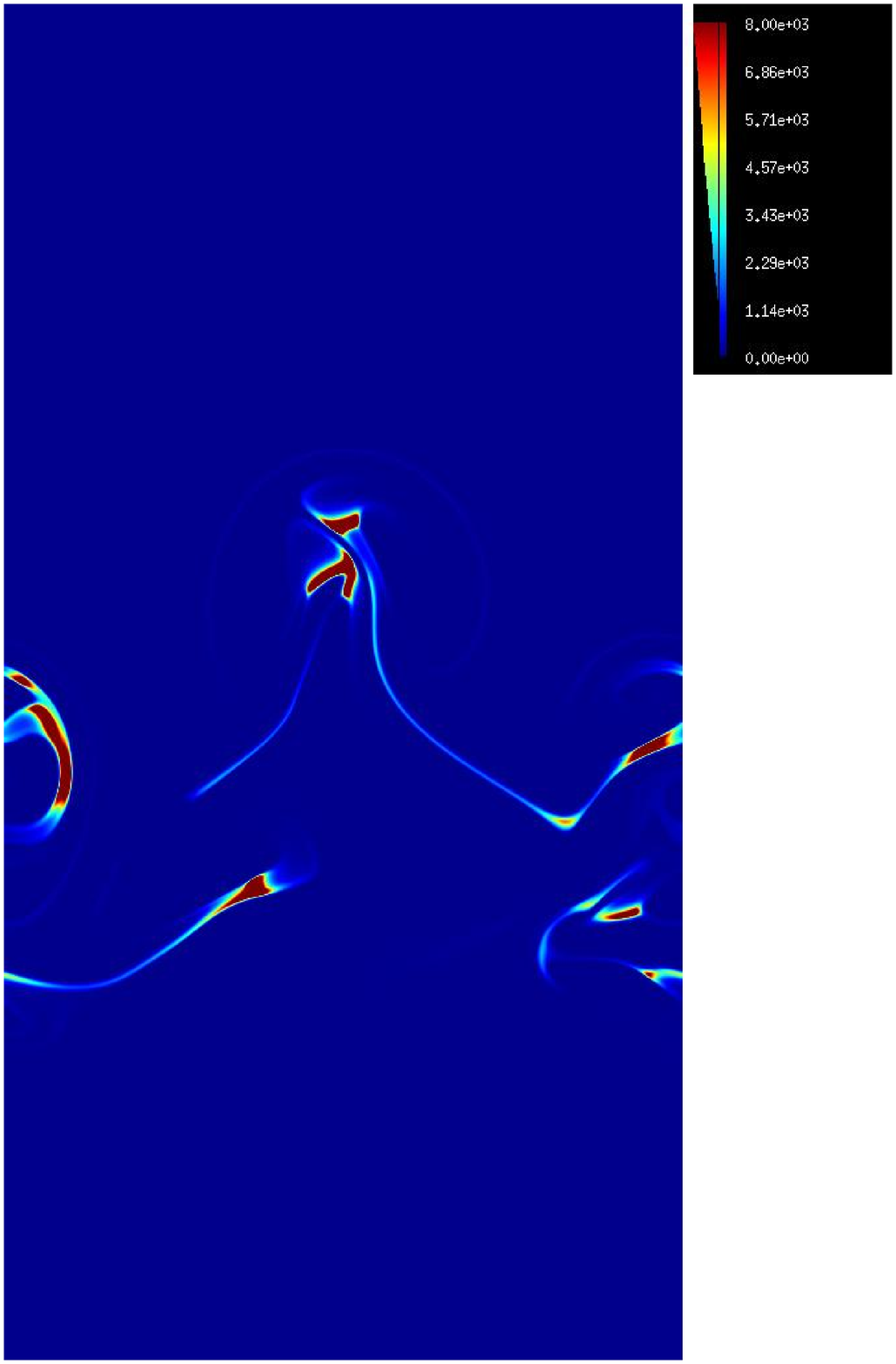}
\epsscale{1.0}
\end{center}
\caption{\label{fig:enuc2d} Absolute value of the carbon destruction
rate for the two-dimensional calculation at $1.16\times 1-^{-3}$~s.
The limits of the colormap are chosen to match the laminar burning
rates, so regions that are deep red are burning faster than laminar.}
\end{figure*}

Figure~\ref{fig:enuc} shows a volume rendering of the magnitude
of the carbon destruction rate at a late time, showing that it is
strongly peaked in small regions.  Slices of carbon destruction rate are shown in
Figure~\ref{fig:enuc_slices}.  There is considerable small-scale
structure present, which perturbs the flame and breaks up the reacting region.
This can be compared to the two-dimensional calculation, as shown in
Figure~\ref{fig:enuc2d}.  The limits of the colormap are chosen to
match the laminar rate, but values over 10~times the laminar burning
rate are present in both RT~calculations.

We pointed out in our previous work \citep{SNrt} that in two-dimensions
the flame speed increases slower than the flame surface area, 
due to the effects of curvature and strain.  From
Figures~\ref{fig:speeds} and \ref{fig:length}, we can see how this
trend differs with dimensionality.  While the flame speeds in two and
three dimensions track each other reasonably well, the relative flame
area change is larger in the three-dimensional run.  This suggests that
in three dimensions, for the
density and length scales considered, the curvature and strain effects
may be much more important in determining the local fuel consumption rate,
since the growth in flame area is much more than
overall increase in flame speed.

As Figure~\ref{fig:3dfull} shows, at late times, energy in the large
scales has cascaded to small scales and creates considerable structure in
the 3D flame.  This effect was is observed in the two-dimensional
results, shown in Figure~\ref{fig:2dview}.  The increased flame surface 
complexity in three dimensions was also reported
in \citet{khokhlov1994}.
%  The comparison of these figures show that
%turbulence plays a role in deforming the flame and that
%two-dimensional simulations cannot capture that.

The power spectrum of the velocity field can be used to characterize
the turbulent fluctuations.  However, because our domain is not
triply-periodic, there is a preferred direction (the
direction of the flame propagation) in which we see a jump in velocity
across the flame.  We note that the velocity field can be written as the sum of a
divergence free term and the gradient of a scalar.  We can further decompose the
latter component:
\begin{equation}
{\bf u} = {\bf u}_d + \nabla \phi + \nabla \psi \enskip .
\end{equation}
Here ${\bf u}_d$ is the divergence free velocity field subject to
${\bf u}_d \cdot {\bf n} = 0$ on the boundaries, $\nabla \phi$ is the
compressible component, while $\nabla \psi$ accommodates the velocity
jump due to the inflow and outflow boundary conditions.
%  Turbulence is comprised
%mainly of a vortical flow, and
%\begin{equation}
%\nabla \times {\bf u} = \nabla \times {\bf u}_d \enskip ,
%\end{equation}
%suggesting that we
Only the divergence-free part of the velocity field is considered when
computing the power spectrum.  We decompose the velocity field using a
projection operation similar to the projection step in the main
hydrodynamics algorithm \citep{SNeCodePaper}.  Since the vorticity
level is negligible at the boundaries, we can treat ${\bf u}_d$ as
though it were periodic in $z$, interpolate it onto a uniform grid at
the finest resolution and decompose the resulting field using a
standard FFT algorithm.  We then define the power spectrum
\begin{equation}
\label{eq:ps}
E(K) = \int_{K=|k|} \mathrm{d}k \; (\hat {u}(k)^2 + \hat {v}(k)^2  + \hat {w}(k)^2 )\,,
\end{equation}
where $\hat u(k)$, $\hat v(k)$, and $\hat w(k)$, are the Fourier
transforms of the $x$, $y$, and~$z$ components of the divergence free
velocity field.  This integral is evaluated by summing the integrand
over spherical shells of constant radius $k = \sqrt{k_x^2 + k_x^2 +
k_z^2}$ and unit wavenumber thickness.

Figure~\ref{fig:3dfull} shows the power spectrum along side the carbon
mass fraction rendering for several different times.  Two curves are
shown in the power spectrum;  the solid line represents the spectrum of the
projected component, while the dashed line represents that of the unprojected field.
%As we see, in the high wavenumbers, these two
%curves are in good agreement.  The discrepancy at low wavenumber is
%due to the jump across the vertical boundaries from the
%inflow and outflow conditions in the non-projected case.
At early time, we see substantial discrepancy in the two curves illustrating the imprint of the 
flame on the spectrum. At later times, this effect dissipates as the energy in the vortical component
increases.
The spikes at
the highest wavenumbers are due to jumps in
refinement in the AMR grid. 
As the RT~instability evolves, the high wavenumbers begin to follow a 
Kolmogorov spectrum, $E(k)\sim k^{-5/3}$, shown as a gray line in the figures.
We can compare the power spectra to the
corresponding carbon mass fraction volume rendering alongside it in
Figure~\ref{fig:3dfull}.  We first achieve the $-5/3$ scaling in
panel~d, and comparing the associated image to the previous ones, we see
that the transition to turbulence is marked by significant wrinkling
on the smallest scales.  Prior to panel~d, the flame surface remains quite
smooth on these small scales due to fire-polishing.

In the final panel, the Kolmogorov turbulence spans about a decade and
a half in spatial scales.  Figure~\ref{fig:spectrum_multi} shows the
final three power spectra (panels~d-f) plotted on the same axes.  We
see that the small scale cutoff to the turbulent cascade does not move
once we are fully turbulent.  We also note that the fire-polishing
length (3.0~cm) corresponds to a wavenumber,~$k_{\mathrm{fp}}$, of 18 (since
$\lambda k = L$, the size of our domain), and the flame thickness
($\delta = 0.55$ to $1.25$ depending on the definition) corresponds to
a wavenumber, $k_{\delta}$, of 98 to~43.  We highlight these in
Figure~\ref{fig:spectrum_multi}.  In these late time power spectra, the
turbulent cascade follows the Kolmogorov scaling well past the
fire-polishing wavelength, proving that the turbulence is affecting
the flame on scales that the~RT alone cannot.  The cutoff occurs
around the flame thickness.

Now that we have determined that the spectrum of turbulence follows
Kolmogorov-like scaling, we can
compute the integral scale and turbulent intensities in each direction
and define a measure of anisotropy.  The turbulent integral scale is defined as
\begin{equation}
\label{eq:integralscale}
l_t^{(x)} =  \left (\frac{1}{ \int_{\Omega} \mathrm{d}\Omega \, u^2} \right )
\int_{\xi=0}^{L_x/2} \mathrm{d}\xi \int_{\Omega} \mathrm{d}\Omega \,\,
u(x,y,z) \, u(x+\xi,y,z) \enskip ,
\end{equation}
\citep{peters:2000}
and is a measure of how correlated the velocity is in $x$-direction.
Here, $u$ is the $x$-component of the velocity, $L_x$ is the physical
size of the domain the $x$-direction, and $\Omega$ represents
the computational domain.  We note that because we use
periodic boundaries, the $\xi$~integral is only over half of the
domain.  A similar definition is used for the $y$ and $z$~integral
scales.  We also define the turbulent intensity,
\begin{equation}
u^{\prime} = \sqrt{\int_{\Omega} \, \mathrm{d}\Omega (u - \bar{u})^2 } \, ,
\end{equation}
where $\bar{u}$ is the average $x$-velocity \citep{peters:2000}.  We
perform all these integrals on the projected velocity field.
Table~\ref{table:integralproperties} lists these parameters for the
six snapshots shown in Figure~\ref{fig:3dfull}. The integral
properties in $x$ and $y$ match well, but in the vertical direction,
$z$, both the integral scale and the turbulent intensities are much
larger; the turbulence is significantly anisotropic.  As an
alternative approach to quantifying the anisotropy, we plot two-point
correlations in each direction as a function length (ie, plot the
magnitude of the $\xi$~integral in eq.~[\ref{eq:integralscale}] as a
function of separation distance).  Figure~\ref{fig:correlation} shows
this quantity separately for each coordinate direction based on the
data plotted in Figure~\ref{fig:3dfull}f.  The two-point correlations
for the $x$ and $y$-directions track each other well, even showing a
strong anti-correlation in the range $8-23$~cm.  In the $z$-direction,
we see the curve is much more broad, demonstrating the larger integral
scale.

\begin{figure*}
\begin{center}
\plottwo{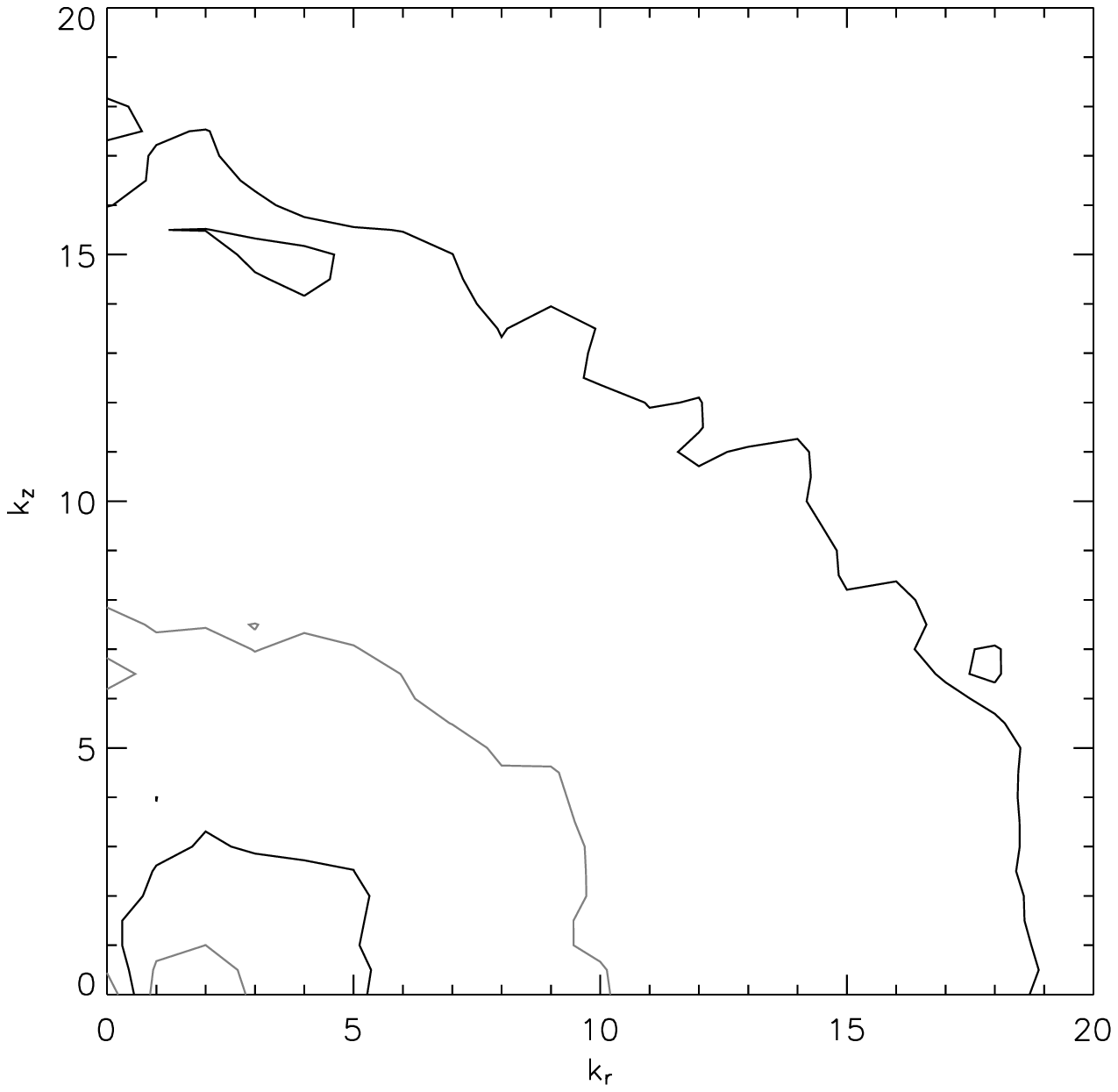}{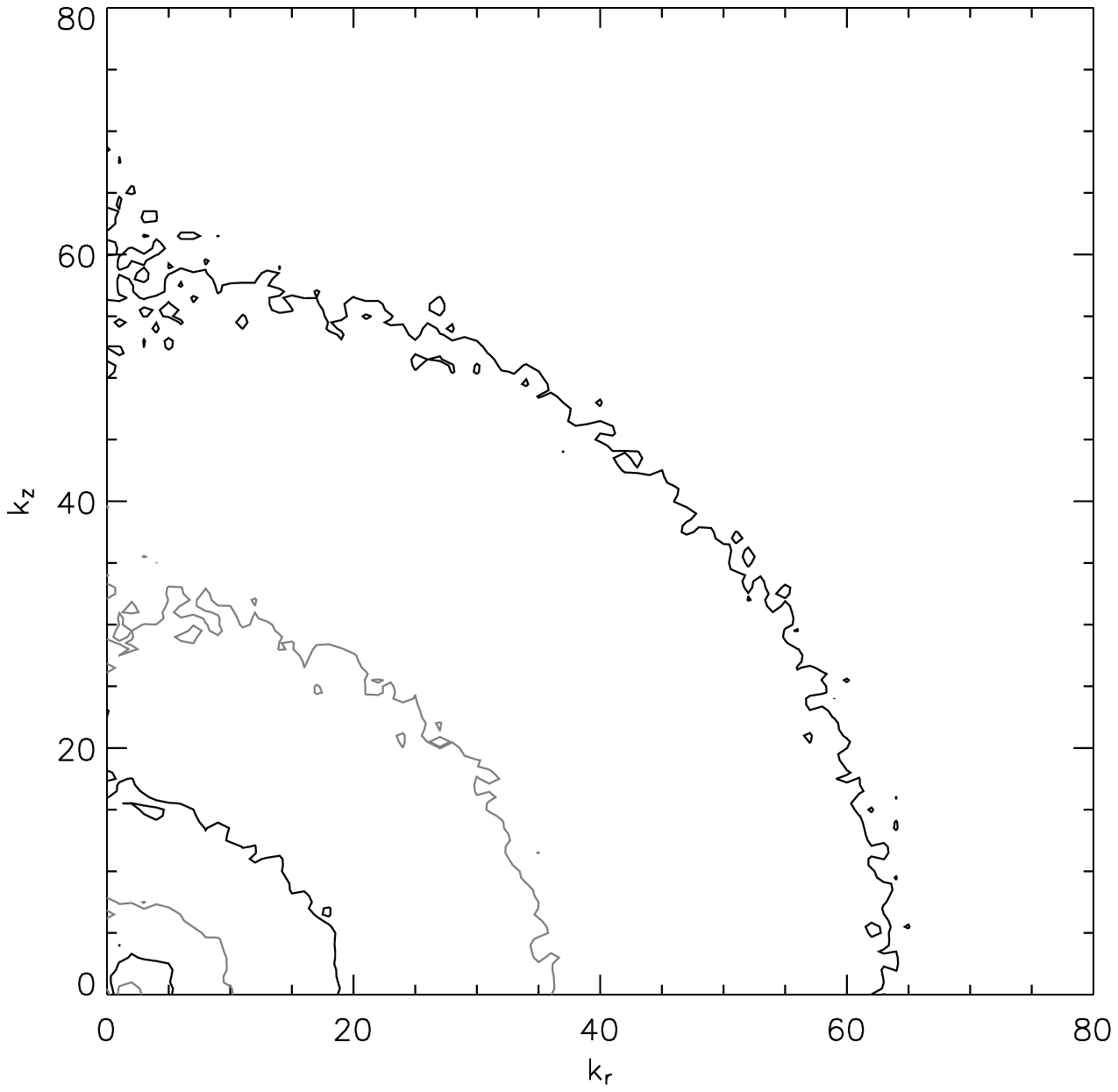}
\end{center}
\caption{\label{fig:anisotropy} Contours of the cylindrically averaged
Fourier transform of the turbulent kinetic energy, $E(k_r, k_z)$ shown
at $E$ = $10^6$, $10^5$, $10^4$, $10^3$, $10^2$, and
$10$~cm$^2$~s$^{-2}$, moving outward from the origin.  The left panel
shows a close up of the origin.  Small wave numbers correspond to
large length scales, and we see that at the very smallest wavenumbers,
we are significantly anisotropic, but as the cascade moves to larger
wavenumbers, we become more and more isotropic.}
\end{figure*}

An interesting question to ask is whether the turbulent field remains
anisotropic on all length scales.  We expect anisotropy on the largest
scales, since gravity is the dominant force term, but as the eddies
interact and cascade to smaller scales, the flow could become more
isotropic.  In the computation of the power spectrum,
Eq.~(\ref{eq:ps}), we integrated over spherical shells of constant
radial wavenumber, thus washing out any information on anisotropy.  We
can look directly at isosurfaces of $E(k_x, k_y, k_z)$ in Fourier
space to see how the degree of anisotropy varies with wavenumber, and
therefore, physical scale. A simplification we can make is to average
over the cylindrical angle in the $k_x$-$k_y$ plane, since we are
isotropic in the direction transverse to gravity.  Then we have
\begin{equation}
E(k_r, k_z) = \left ( \frac{1}{\int_{k_r = |k^{\prime}|} \mathrm{d}k}
\right ) \int_{k_r = |k^{\prime}|} \mathrm{d}k \; E(k_x, k_y, k_z)
\enskip ,
\end{equation}
where $k^{\prime} = \sqrt{k_x^2 + k_y^2}$.  If the turbulent kinetic
energy were isotropic, we would expect contours of constant $E(k_r,
k_z)$ to be circles in the $k_r$-$k_z$ plane, and the largest scales
in the flow would correspond to circles of smaller radius.  
Figure~\ref{fig:anisotropy} shows these contours for several values of
$E(k_r, k_z)$, every decade from 10 to $10^6$~cm$^2$~s$^{-2}$.
In the left panel, we look at the smallest
wavenumbers.  We see that the innermost curves are not at all
circular---demonstrating that at the largest scales, we are quite
anisotropic. This supports the findings from the integral scale and
turbulent intensity comparison.  As we go to higher
wavenumbers, the contours become more and more circular, demonstrating
that the turbulence becomes more isotropic as it cascades down to
smaller scales.  The outermost contour matches a radial wavenumber of
$\sim 60$, which is close to the highest wavenumber still within our
turbulent cascade.  We conclude that once the turbulence has cascaded
down through about an order of magnitude of length scales, it is
practically isotropic.

We see that our laminar flame speed is much smaller than the turbulent
velocities at late times.  In previous work that looked at the scaling
of the turbulent flame speed with area enhancement, the smallest ratio
of laminar speed to turbulent velocity considered was $S_l/w^{\prime}
= 0.95$ \citep{nbr1999}, at which they found reasonable scaling of the
turbulent flame speed to the area enhancement.
Here, we reach a point where $S_l/w^{\prime} = 0.3$, and the
dominance of the turbulent velocity over the laminar flame speed in
our calculation results in a large departure from perfect scaling of
the turbulent flame speed to the area enhancement.

A common estimate of the turbulent velocity is to apply the
Sharp-Wheeler scaling for non-reactive RT:
\begin{equation}
\label{eq:rtvel}
v_{\mathrm{RT}} \sim \frac{1}{2} \sqrt{\mathrm{At}\, g L} 
\end{equation}
\citep{sharp1984,daviestaylor1950}, where $\mathrm{At}$ is the Atwood number
and $L$ is the integral scale of the turbulence.  This is the
magnitude of the velocity due to RT~growth.  If we consider the $z$
direction only---since this is the direction where RT is doing the
most work, we find $v_{\mathrm{RT}} = 2.7\times
10^4~\mathrm{cm~s^{-1}}$ for case~f, just slightly above the measured
$w^{\prime}$.  This supports the idea that the turbulence cascade is
the result of the RT~instability.  It is unclear what the analog of
equation~(\ref{eq:rtvel}) should be for the transverse directions.

In turbulent combustion, the turbulence will affect the flame on all
scales down to the one where the flame can burn away an eddy before
its turnover time.  This scale, the Gibson scale, represents the
smallest turbulent scale that can influence the flame, and together
with the flame thickness and fire-polishing length are the minimum
scales required to resolve on our grid without including explicit
subgrid models for turbulence of the flame.  The Gibson scale is
\begin{equation}
\label{eq:gibson}
l_G = l_t \left ( \frac{S_l}{u^{\prime}} \right )^3 
\end{equation}
\citep{peters:2000} for Kolmogorov turbulence (although,
\citealt{khokhlov1995} argues the true scale is a factor of $\sim 500$
larger because of flame speed enhancements due to cusping).  Again,
due to the anisotropy, this can vary with direction, but the result
falls between $0.38$ and $0.44$~cm using the data from
Table~\ref{table:integralproperties}.  We note that this is very close
to our flame thickness, 0.55~cm, and about a 4~zone resolution, so it
should just be resolved on our grid.  Although we note that since we
are not using an explicit viscosity in our calculation, we would
expect the cascade to cutoff here due to numerical viscosity as well.

Figure~\ref{fig:borghi} shows a combustion diagram
\citep{poinsotveynante,peters:2000}, used in terrestrial combustion
for determining the flame regime.
Three sets of points are provided, using the $x$, $y$, and $z$
integral scales and turbulent intensities separately.  The vertical
axis is the turbulent intensity divided by the laminar flame speed and
the horizontal axis is the integral scale divided by the flame
thickness.  The different combustion regimes are separated by lines
of unit Karlovitz number (the ratio of the nuclear reaction timescale
to the Kolmogorov time), unit Damk\"ohler number (the ratio of the
largest eddy turnover time to the nuclear reaction time scale) and
forces at the integral scale) \citep{poinsotveynante}.  
The flame
begins in the lower-left region of the diagram and evolves toward the
top-right with time.  The points corresponding to the $z$-based
integral quantities are always more to the upper right of the
corresponding $x$- and $y$-based points.  The final points appear to
cross the line of constant Karlovitz number, suggesting that
in three dimensions, we enter the distributed burning regime at this
density.  This is supported by our measurement of the Gibson scale,
being just below the flame thickness.  We note that we used
equation~(\ref{eq:ft1}) for the flame thickness, and because of this
ambiguity in the definition, these points would shift to the
upper-left using equation~(\ref{eq:ft2}).

% ============================================================================
%  Conclusions
% ============================================================================
\section{CONCLUSIONS}
\label{sec:conclusions}

\begin{figure*}
\begin{center}
\epsscale{1.0}
\plotone{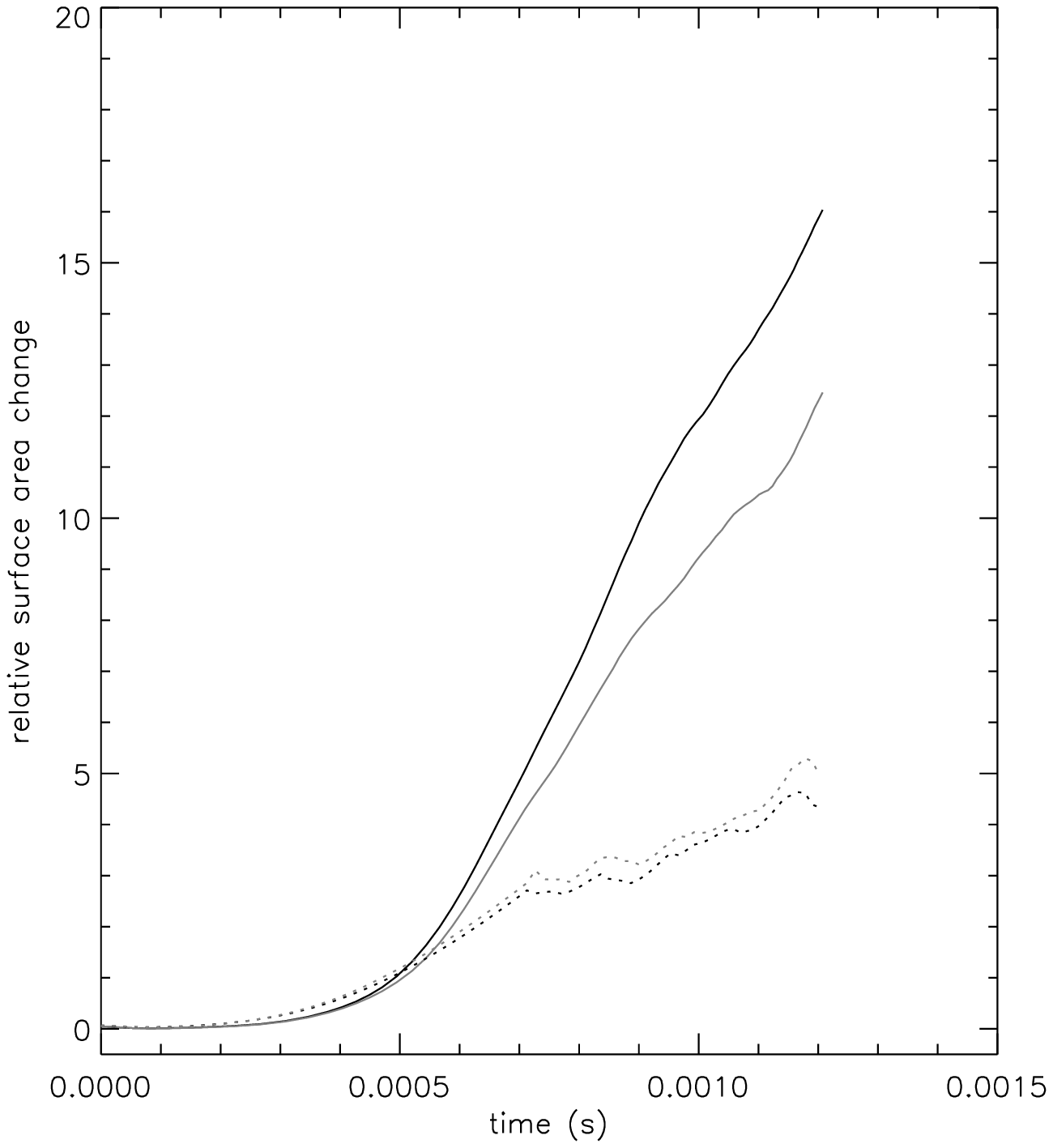}
\epsscale{1.0}
\end{center}
\caption{\label{fig:length} Relative surface area change for the
two-dimensional (dotted) and three-dimensional (solid) flame
simulations.  Two curves are shown for each dimensionality,
corresponding to a carbon mass fraction of 0.25 (black) and 0.1
(gray).  In three dimensions, the relative change in surface area is
far larger than the two-dimensional case, but the flame speed changes
are about the same (Figure~\ref{fig:speeds}), demonstrating the larger
effect of curvature and strain on the local burning rate.}
\end{figure*}

We presented a three-dimensional simulation of an RT~unstable flame in
SNe~Ia, following the evolution into the fully turbulent regime---the
first such calculations performed without a flame model.  We showed
that the growth of the mixed region is slightly faster in two
dimensions than in three, but at late times, turbulence takes over in
the three-dimensional case, providing a large late-time acceleration
of fuel consumption.  Curvature and strain effects apparently play a
larger role in the three-dimensional simulations, since the relative
surface area increases much more in the three-dimensional case than in
the two-dimensional case (Figure~\ref{fig:length}), for roughly the
same increase in flame speed over the course of the calculation.  Any
model of burning in SNe~Ia should therefore be based on
three-dimensional results, in agreement with conclusions presented in
\citet{khokhlov1994}.

The turbulence is Kolmogorov in nature as we showed, with the energy
spectrum at late times following a $-5/3$ power law scaling in
wavenumber for over one and a half orders of magnitude.  Apparently
the previously postulated Bolgiano-Obukhov statistics
\citep{niemeyerkerstein1997} are not relevant to this Type~Ia
supernova regime, or perhaps apply only to data from two dimensional
\citep{chertkov2003} models.  We also demonstrated that our turbulent
cascade reached the smallest scales permitted by the resolution
capabilities of our numerical algorithm on this grid.  The turbulent
intensities are consistent with the velocities expected from the pure
RT~instability (eq.~[\ref{eq:rtvel}]), suggesting that it is being
driven by the large scale RT~growth.  In a real star, there will be
even larger scales that are RT~unstable, suggesting that a `supergrid'
model should be applied to force our domain with the expected
intensities.  This does not change the results presented here 
qualitatively, but the addition of externally fed turbulence would
scale the numbers presented in Table~\ref{table:integralproperties}.
This is a separate problem and could be the focus of future
simulations.

The turbulence is strongly anisotropic on the largest scales---with
the integral length scale in the vertical direction more than five
times that in the transverse directions.  But as we showed, the
turbulence becomes more isotropic at smaller scales.
%Large scale calculations, can deal with three orders
%of magnitude of length scales, with modern, adaptive mesh
%techniques. \MarginPar{new text here}
The inertial range of the turbulence in full star calculations is less
than two orders of magnitude, since at the largest scale, stellar
flows are not turbulent and the computing hardware and computing
algorithm sets a minimum resolvable scale.  At the very largest
scales, gravity drives anisotropic fluctuations.  Based on our
measurements of the anisotropy, a minimum of a decade of length scales
must be properly represented on the grid in order to permit valid use
of an isotropic subgrid model.  Isotropy on the grid scale
is currently employed in turbulent flame subgrid
models, notably \citep{niemeyerhillebrandt1995b,reinecke2002b}, and
our findings verify that this is a good assumption, provided that a
sufficiently large interval of the turbulence spectrum is accurately
simulated.  Isotropic subgrid scale turbulence would predict isotropic
turbulent flame speeds and lead to regular, somewhat spherical
expansion of any ignition points.  The situation would be quite
different if anisotropy of the turbulence were important on the small
scales.  We would then expect the burning in the direction of gravity
to greatly out-pace the lateral burning.  This may mean that more
ignition points would be required in an anisotropic turbulence-based
simulation to match the characteristics of an isotropic
turbulence-based simulation.

Previously, the differences between two- and three-dimensional
RT~flames
were explored by \citet{khokhlov1994,khokhlov1995}.  These
calculations employed a thick-flame model, where the flame thickness
was stretched to be resolvable on the grid.  These thickened-flames
did not, however, employ a subgrid model, so they neglected the
interactions between the flame and the turbulence on scales between
the real flame thickness and its thickened counterpart.  The present
calculations resolve the fire-polishing length, the flame
thickness, and the Gibson scale.  

The transition to distributed burning occurs when wrinkling occurs on
scales smaller than the flame thickness.  There are two dominant
mechanisms for this wrinkling in Type~Ia supernovae---turbulence and
the RT~instability.  In our two-dimensional study, only the
RT~instability influenced the flame, because of the nature of
two-dimensional turbulence, so the transition to distributed burning
occurs at the density where the fire-polishing length becomes smaller
than the flame thickness.  As we see in the present study,
the three-dimensional turbulent cascade can reach scales smaller than
the fire-polishing length and greatly wrinkle the flame on scales not
possible in two dimensions.  This wrinkling will occur on scales down
to the Gibson length, where the small eddies can then burn away before
they can perturb the flame.  We presented a combustion diagram, common
in studies of terrestrial flames, which suggests that at this
density, the simulation evolves into the distributed burning regime.

\begin{figure*}
\begin{center}
\epsscale{1.0}
\plotone{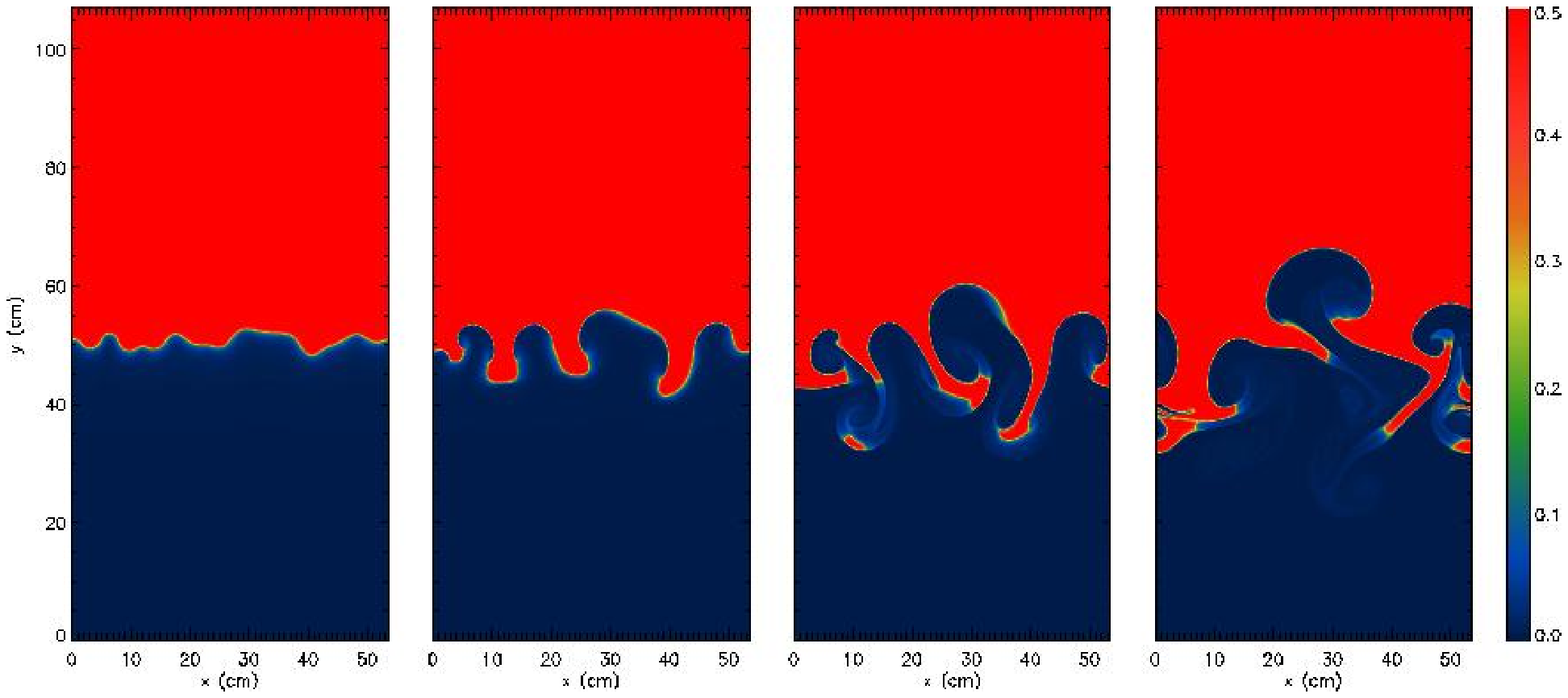}
\epsscale{1.0}
\end{center}
\caption{\label{fig:2dview} Carbon mass fraction for the
two-dimensional simulation at $2.5\times 10^{-4}$, $5\times 10^{-4}$,
$7.5\times 10^{-4}$, and $10^{-3}$~s.  In two dimensions, the
turbulence does not reach the small scales, resulting in a much
smoother flame.}
\end{figure*}

\cite{khokhlov1994} concluded that more small scale structure was
present in the three-dimensional simulations, due to the nature of the
turbulent cascade in three dimensions.  We show overwhelming support
for this in our simulations, as shown in Figures~\ref{fig:3dfull} and
\ref{fig:2dview}.  \citet{khokhlov1994} argues that this difference is
most important when the turbulence from the nonlinear
RT~cascades to higher wavenumber (as it only can do in three dimensions) and reaches
scales smaller than the fire-polishing length, and that it is this
small scale turbulence that further wrinkles the flame thereby increasing the
energy generation rate.  With a grid that is effectively $512\times
512\times 1024$ zones, we expect that we are modeling the turbulence
quite well.  At late times, we see a $-5/3$ power law for the turbulent
spectrum, in agreement with Kolmogorov theory.  The two-
and three-dimensional integral quantities (flame speed, surface area, and
width) track each other well up to the onset of turbulence.

In our simulations, we never reach a point where the flame speed
reaches a steady state, as reported in \citet{khokhlov1995}---nor do
we expect to.  The only time such an effect was seen was in the
narrowest run of our distributed regime flame in our two-dimensional
study \citep{SNrt}, where we simply run out of modes to grow in our
box---the instability saturates.  In the real problem, this
arrangement will not occur.  The burning rate along the surface of the
flame varies tremendously due to curvature and strain, reaching up to
$12\times$ the laminar value.  It is simply not the case, for our
simulation, that the overall effective flame speed is the surface area
enhancement multiplied by the laminar flame speed.

\begin{figure*}
\begin{center}
\epsscale{1.0}
\plotone{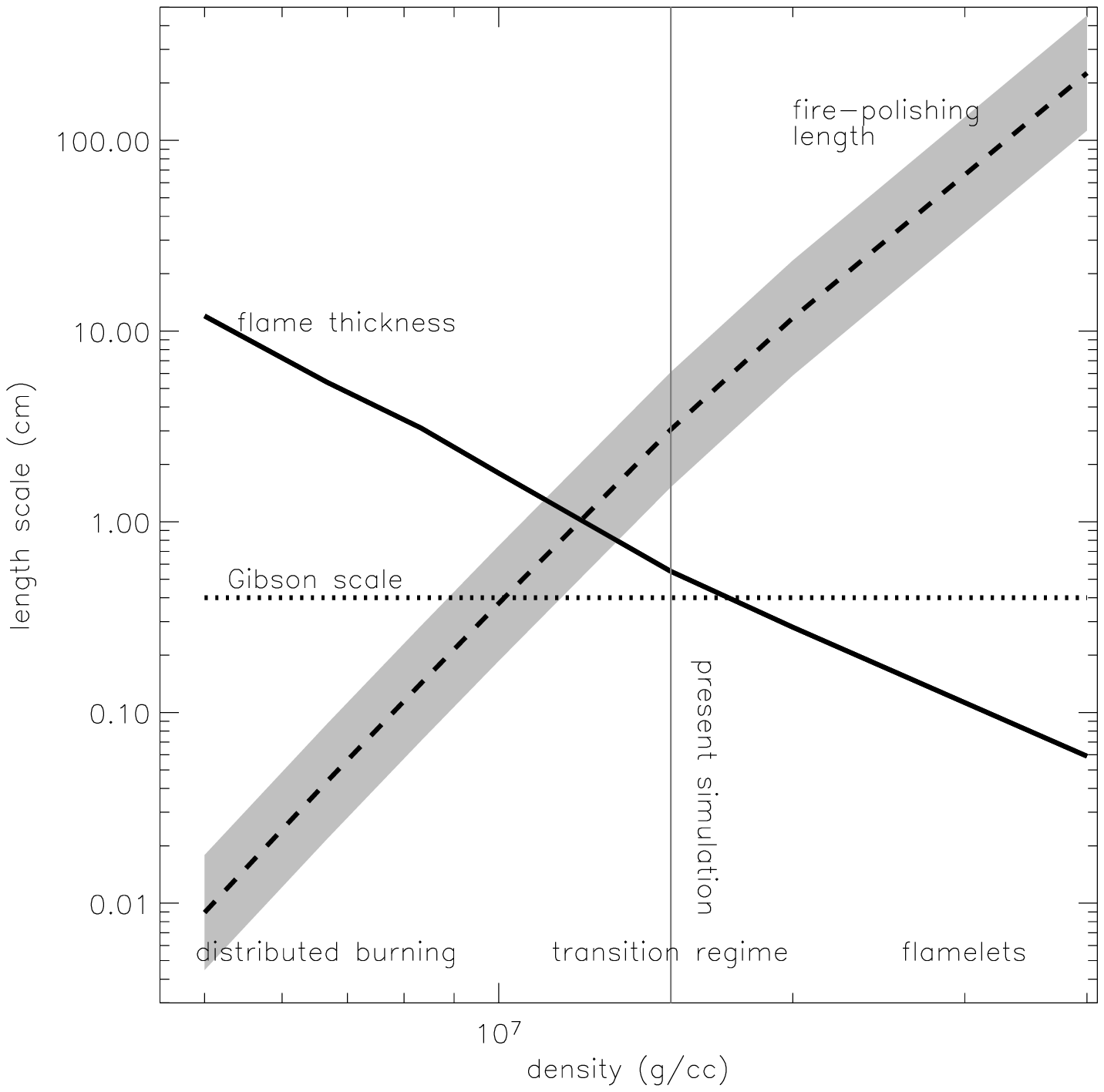}
\epsscale{1.0}
\end{center}
\caption{\label{fig:regimes} Flame thickness (solid), fire-polishing
and length (dashed) as a function of density.  The shaded region
reflects a factor of two uncertainty in the fire-polishing length, due
either to changing gravitation acceleration or uncertainty in the
multiplicative constant.  We see the fire-polishing length and flame
thickness lines crossing at a density between $10^7~\gcc$ and
$1.5\times 10^7~\gcc$.  The Gibson scale line (dotted) is at the
measured value for the present three-dimensional simulation (see
\S~\ref{sec:results}), provided for comparison.  At the high density
side, we are in the flamelet regime.  At low densities, we are in the
distributed burning regime.}
\end{figure*}

Our calculations were performed at a density of $1.5\times 10^7~\gcc$.
Because of our requirement to
resolve the flame thickness while fitting the fire-polishing length
and Gibson scale on the grid (as shown in Figure~\ref{fig:regimes}),
we are unable to perform simulations at the $10^8~\gcc$ density 
considered previously by \citet{khokhlov1994,khokhlov1995}.
We found that the Gibson scale is $\sim 0.4$~cm, very close to the 1.0~cm
estimate by \citet{niemeyerwoosley1997}.  Therefore, our flame is in a
regime where the fire-polishing length, Gibson scale, and flame
thickness are all of the same order of magnitude.  The calculations
presented by \citet{khokhlov1994,khokhlov1995} at a density of
$10^8~\gcc$ have a true flame thickness of $2.75\times 10^{-2}$~cm
\citep{timmeswoosley1992}, which is much smaller than estimates of the
Gibson scale at that density \citep{niemeyerwoosley1997} or the
fire-polishing length ($5\times 10^4$~cm, according to
\citealt{khokhlov1994}).  The thickened flame thickness is several
grid cells in his calculation, or $\sim 10^5$~cm, which is much larger
than the Gibson scale, and of the same order as the fire-polishing
length.  Since the flame-thickening used in these calculations does
not account for the turbulence interactions on scales between the true
flame thickness and the Gibson scale (such as through an efficiency
function as in \citealt{colin2000}), it is likely that much of the
actual burning is neglected.  The optimization of an efficiency
function approach for correcting thick-flame RT~calculations
will be the subject of a future study.

There are also differences in the range of length scales captured in
the simulations.  The grid employed in \citet{khokhlov1995} ranged
from~32 to 85~zones across and~256 to~512 in the vertical direction.
The inertial range representable on this grid is quite small; it is
unlikely that fully developed turbulence was present.  Furthermore,
the grid resolution in most cases was only $4\times$ the
fire-polishing length, so it is unclear how much the interaction
between turbulence on scales smaller than the fire-polishing length
and the flame could have influenced those calculations.  Our grid is
512~zones across and almost $30\times$ the fire-polishing length, and
as we show, we have a fully developed turbulence on scales between the
fire-polishing length and the grid resolution.  Finally, we seed more
modes, about 10 in each direction compared to 3 in
\citet{khokhlov1994,khokhlov1995}, which will allow for a longer
duration of the non-linear phase of the RT~instability.

In the flamelet regime, there should exist a
scaling to an arbitrary size domain for regimes having the same
reactive Froude number,
\begin{equation}
\mathrm{Fr} = \frac{S_l}{\sqrt{\mathrm{At}\, g L_0}} \sim
\sqrt{\frac{\lambda_{\mathrm{fp}}}{L_0}}
\end{equation}
which evaluates the relative importance of the laminar flame speed to
gravity.  We expect this scaling to hold true in the high density case of
our two-dimensional studies \cite{SNrt} and in the early time
evolution of the present study.  The present 3D simulation is at
$\mathrm{Fr} = 7 \times 10^{-2}$, and since $(\mathrm{At}\,g)$ is roughly
constant throughout the whole explosion, this suggests that our
calculation would be very similar to another calculation, e.g., on a
length scale of 50~km (1000 times larger than our present one) for a
timescale 30 times longer (since $U \sim \sqrt{L}$), up until the
point where the present calculation just begins to enter the
distributed regime.
More precisely, at a larger scale the front
geometry would be the same for the range of length scales we can model; however,
the flame would not be fully resolved and additional subgrid structure would be
present that could potentially modify the overall flame dynamics.
%Obviously at such a scale, the flame would not be
%resolved and subgrid structure would be present, but the front
%geometry for the range of length scales we can model should be the
%same.
A comparison of thickened flames at the same Froude number of
the present calculation would be an important validation of that flame
model.  Once validated, our approach offers a potential path toward
bootstrapping to studies that include the entire star. The next
is to consider turbulence-flame interactions directly,
by inflowing turbulent fuel into the domain.  In particular, it should
be possible to develop a turbulent flame model that allows for the
present simulation to be repeated with much less resolution, which
retaining the overall integral properties.

The results of our three-dimensional resolved study of the reactive RT
instability are encouraging.  We have shown that we can accurately
model a fully turbulent flame in three dimensions.  At late times, as
our combustion regimes diagram shows, we just begin to enter the
distributed burning regime.  There is still a lot of speculation that
a transition to detonation can occur in this regime, but our
two-dimensional studies \citep{SNrt} suggest that this could not
occur.  A followup to those studies, well in the distributed regime
will be the focus of a later study.

\acknowledgements 

We thank Jens Niemeyer, Alan Kerstein, and Jonathan Dursi for useful
discussions regarding the turbulence and flame interactions, and Frank
Timmes for making his equation of state and conductivity routines
available online.  Support for this work was provided by the DOE grant
No.\ DE-FC02-01ER41176 to the Supernova Science Center/UCSC and the
Applied Mathematics Program of the DOE Office of Mathematics,
Information, and Computational Sciences under the U.S. Department of
Energy under contract No.\ DE-AC03-76SF00098.  SEW acknowledges NASA
Theory Award NAG5-12036.  The first part of this calculation was
performed on the IBM SP (seaborg) at the National Energy Research
Scientific Computing Center, which is supported by the Office of
Science of the DOE under Contract No.\ DE-AC03-76SF00098.  This
calculation was finished on the Columbia system at NASA Ames Research
Center.  We are grateful to both centers for the computer time.
Movies of the simulations are available at
\url{http://www.ucolick.org/\~{}zingale/rt3d/}.

\vskip 5 mm

\clearpage

%\bibliographystyle{apj}
%\bibliography{rt}

% ============================================================================
%  Figures
% ============================================================================

%\begin{figure*}
%\begin{center}
%\epsscale{0.6}
%\plotone{rt3d_full.eps}
%\epsscale{1.0}
%\end{center}
%\caption{\label{fig:3dfull} Carbon mass fraction at $5\times
%10^{-4}$~s (top) and $10^{-3}$~s (bottom) showing the early
%development of the RT instability and the late time transition to
%turbulence.}
%\end{figure*}

\begin{figure*}
\begin{center}
\epsscale{1.0}
\plotone{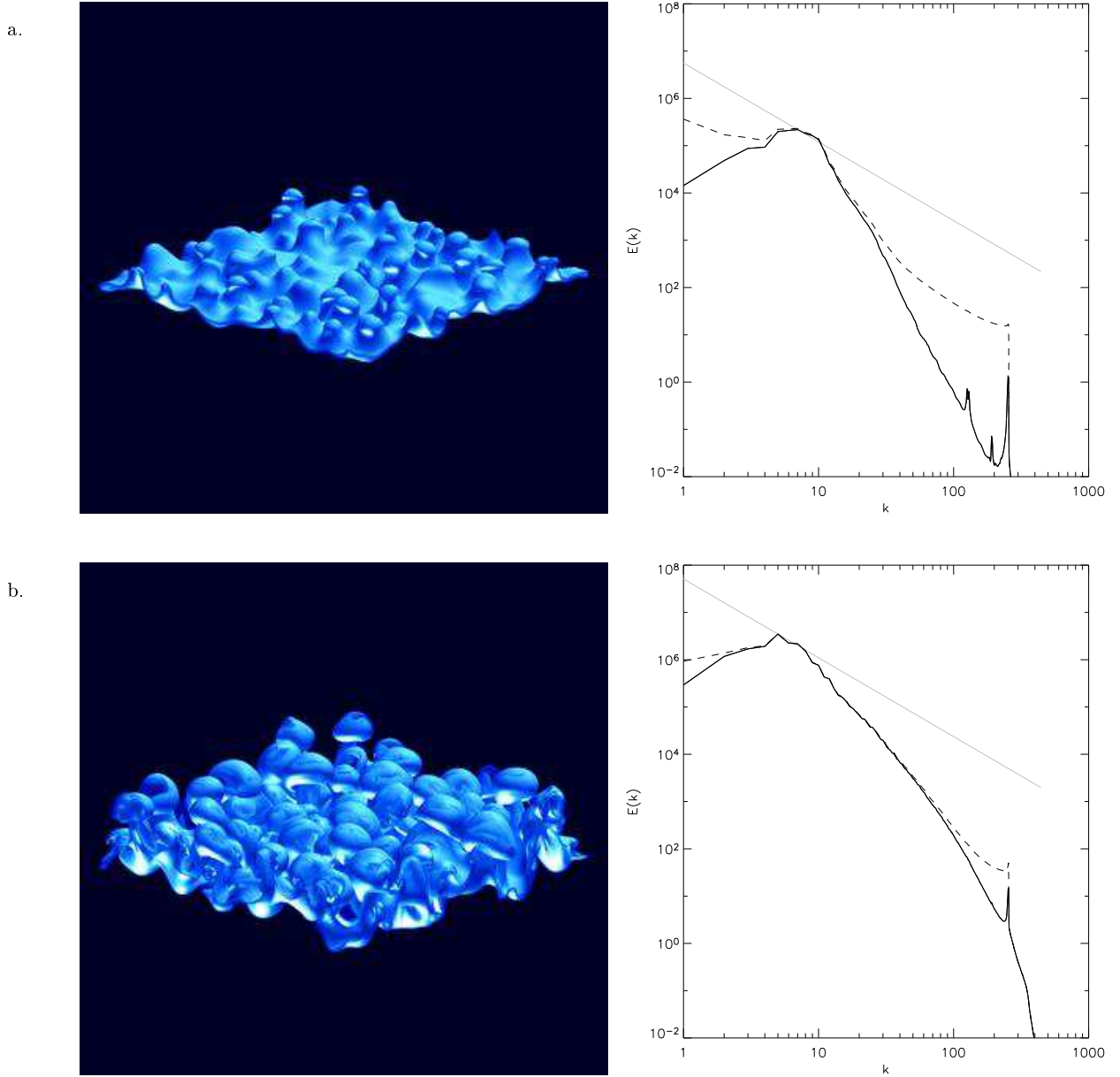}
\epsscale{1.0}
\end{center}
\caption{\label{fig:3dfull} Volume rendering of the carbon mass
fraction (left) and the kinetic energy power spectrum (right) at
(a)~$4.04\times 10^{-4}$~s,
(b)~$6.62\times 10^{-4}$~s,
(c)~$8.11\times 10^{-4}$~s,
(d)~$9.43\times 10^{-4}$~s,
(e)~$1.07\times 10^{-3}$~s, and
(f)~$1.16\times 10^{-3}$~s.  The dark solid curve is the power
spectrum after projecting out the compressible components.  The dashed
curve is without the projection.  The gray line is a $-5/3$ power law.
We see that at late times, we reach a Kolmogorov scaling. }
\end{figure*}

\begin{figure*}
\figurenum{\ref{fig:3dfull}}

\begin{center}
\epsscale{1.0}
\plotone{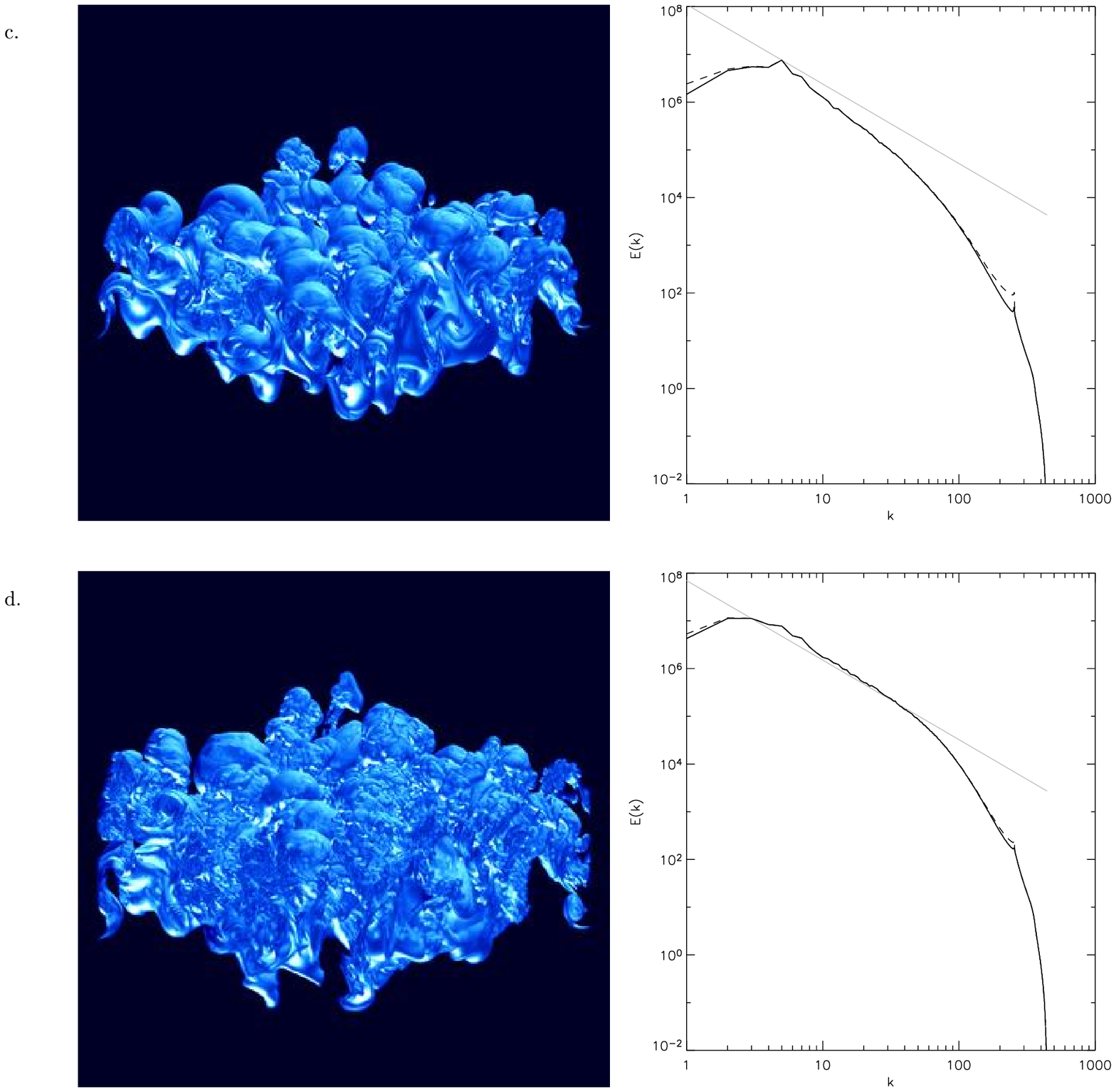}
\epsscale{1.0}
\end{center}
\caption{cont.}
\end{figure*}

\begin{figure*}
\figurenum{\ref{fig:3dfull}}

\begin{center}
\epsscale{1.0}
\plotone{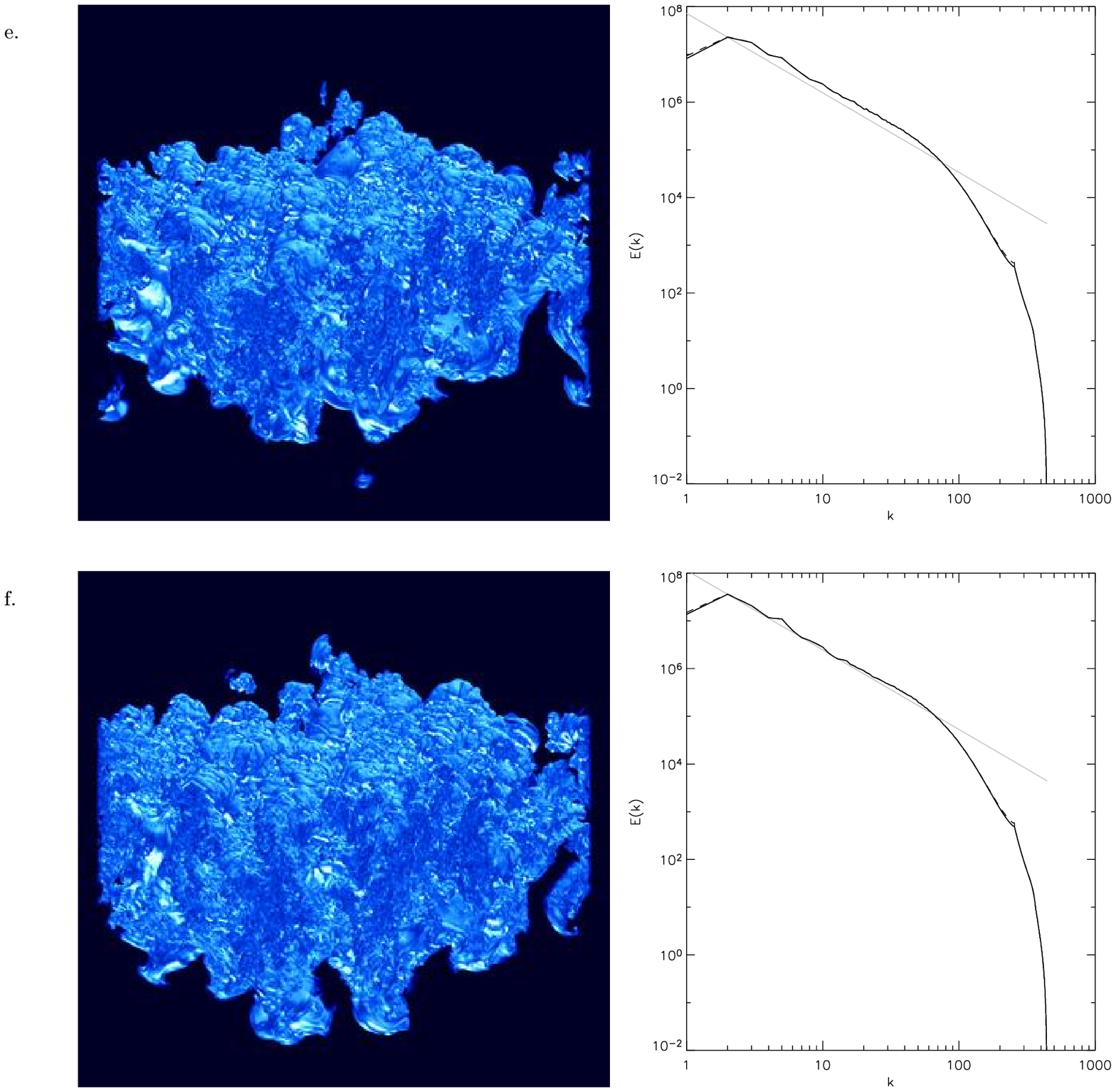}
\epsscale{1.0}
\end{center}
\caption{cont.}
\end{figure*}

\clearpage

\clearpage

%\begin{figure*}
%\begin{center}
%\plotone{panel.ps}
%\end{center}
%\caption{\label{fig:3dview} A time sequence of a small region of the
%three dimensional domain showing the growth of the RT instability and
%the breakdown of the bubbles and transition to turbulence at late
%times.  This region is $160\times 192\times 428$ zones out of the full
%$512\times 512\times 1024$ zone domain.  {\color{red} We don't know
%exactly what times these are from, since they are just taken from the
%animation.  Should we regenerate this panel at specific times
%(i.e. every $2.5\times 10^{-4}$~s as in the 2d panel?}}
%\end{figure*}

%\clearpage

\begin{figure*}
\begin{center}
\plotone{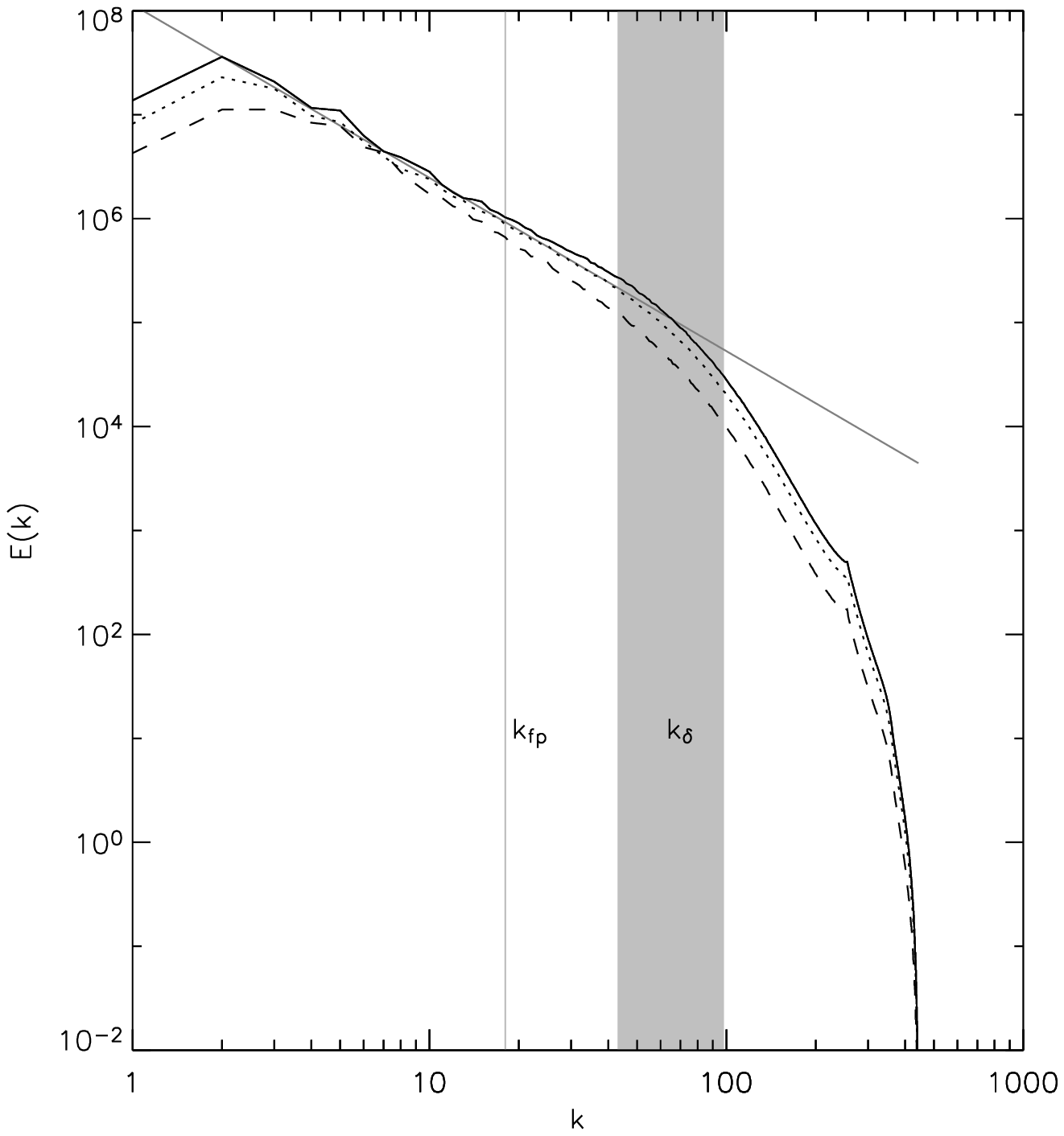}
\end{center}
\caption{\label{fig:spectrum_multi} Late time turbulence power spectra
corresponding to panel~d (black solid line), e (dotted line), and f
(dashed line) of Figure~\ref{fig:3dfull}.  The straight gray line is a
$-5/3$ power law.  Here we see that the small scale cutoff to the
Kolmogorov spectrum does not shift with time with this fully developed
turbulence.  The position of the fire-polishing wavenumber, $k_{\mathrm{fp}}$
and the range of flame thickness spanned by our two definitions,
$k_{\delta}$, are shown as well.}
\end{figure*}

\clearpage

\begin{figure*}
\begin{center}
\plotone{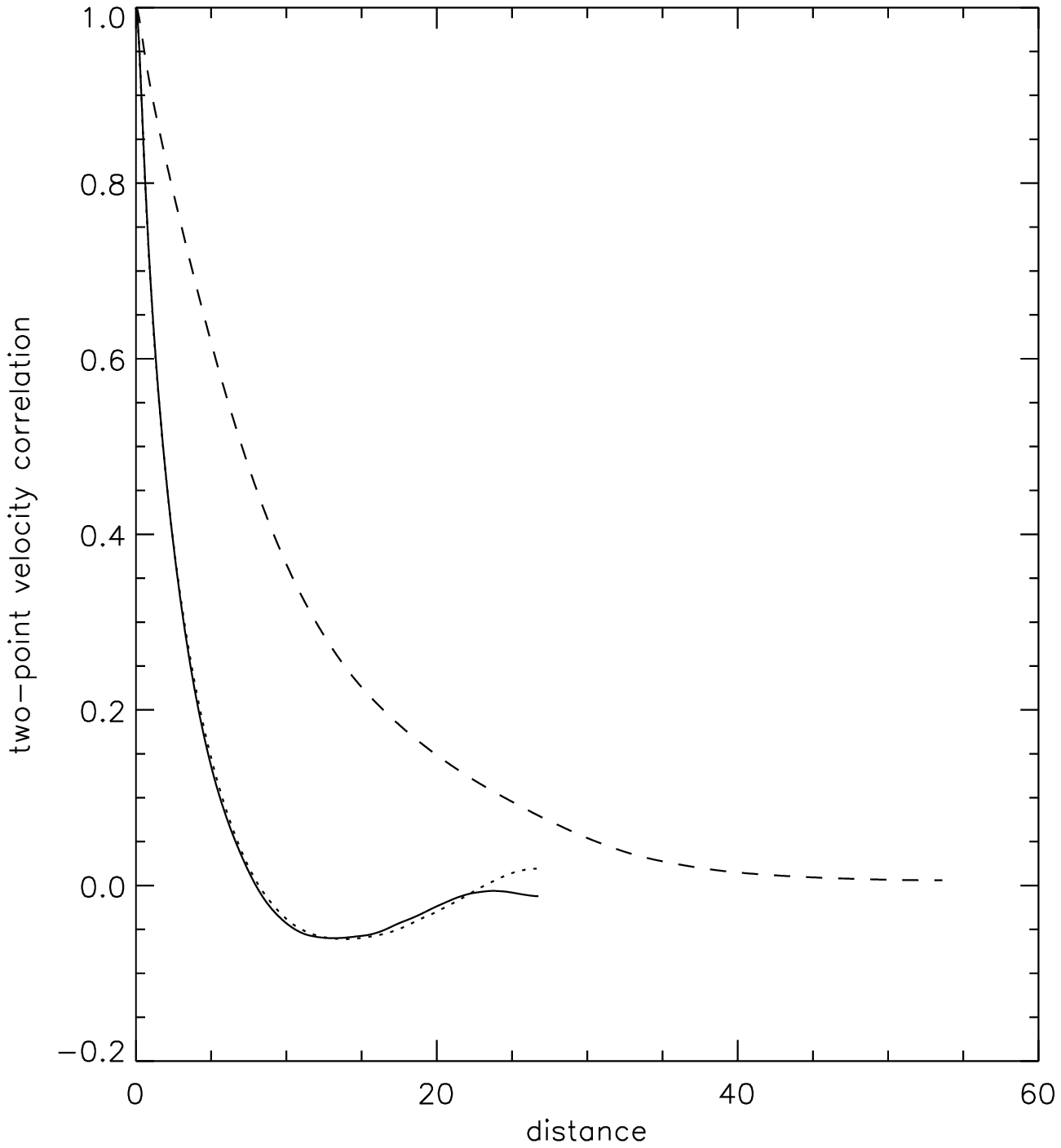}
\end{center}
\caption{\label{fig:correlation} Two-point correlation for data in
Figure~\ref{fig:3dfull}f in each coordinate direction: $x$ (solid), $y$
(dotted), and $z$ (dashed).  We see that in the $z$-direction, the
correlation length scale is much higher than in the transverse
directions.  In the transverse directions, the data is anti-correlated
between 8 and 23 cm.}
\end{figure*}

\clearpage

\clearpage

\begin{figure*}
\begin{center}
\plotone{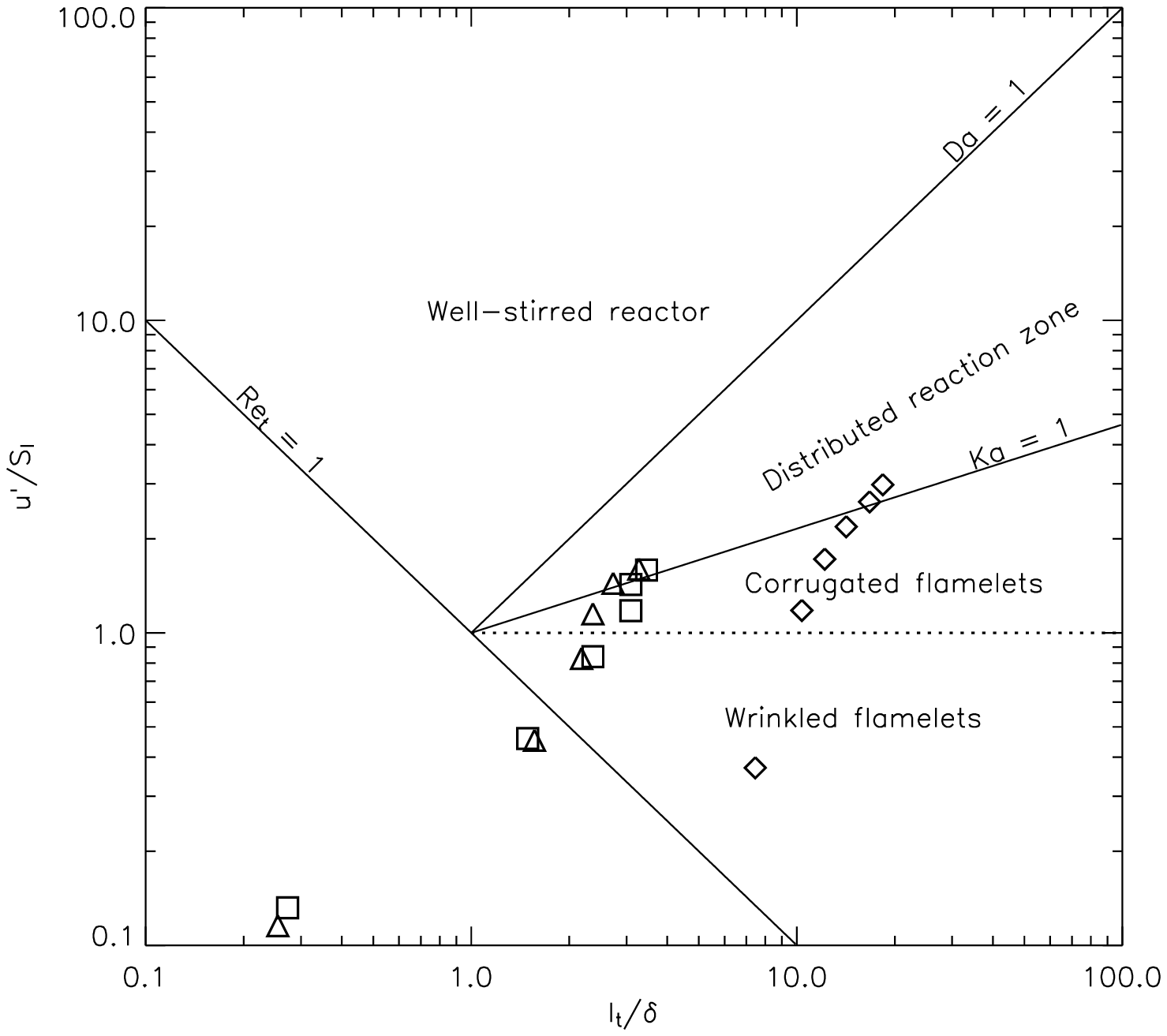}
\end{center}
\caption{\label{fig:borghi} Combustion regimes diagram
\citep{poinsotveynante} using the turbulent intensity and integral
scale for the six timesteps shown in Figure~\ref{fig:3dfull}.  Because
our turbulence is so anisotropic, we plot each timestep three times,
once using the $x$ data (triangles), the $y$ data (squares), and the
$z$ data.  In all cases, the flame moves to the top-right as it
evolves, just passing into the distributed regime at the end of the
calculation. }
\end{figure*}

\clearpage

\begin{table*}
\begin{center}
\caption{\label{table:integralproperties} Integral length scale and turbulent intensities.}
\begin{tabular}{rrrrrrrr}
\tableline
\tableline
\multicolumn{1}{c}{key\tablenotemark{a}}&\multicolumn{1}{c}{time} & \multicolumn{1}{c}{$l_t^{(x)}$} & \multicolumn{1}{c}{$l_t^{(y)}$} & \multicolumn{1}{c}{$l_t^{(z)}$} & 
  \multicolumn{1}{c}{$u^{\prime}$} & \multicolumn{1}{c}{$v^{\prime}$} & \multicolumn{1}{c}{$w^{\prime}$} \\
& \multicolumn{1}{c}{(s)} & \multicolumn{1}{c}{(cm)} & \multicolumn{1}{c}{(cm)} & \multicolumn{1}{c}{(cm)} & \multicolumn{1}{c}{(cm s$^{-1}$)} & \multicolumn{1}{c}{(cm s$^{-1}$)} & \multicolumn{1}{c}{(cm s$^{-1}$)} \\
\tableline
a & $4.04\times 10^{-4}$ & 0.14 &  0.15 &   4.1  &  $9.48\times 10^2$ & $1.08\times 10^3$ & $3.03\times 10^3$ \\
b & $6.62\times 10^{-4}$ & 0.86 &  0.82 &   5.7  &  $3.71\times 10^3$ & $3.76\times 10^3$ & $9.66\times 10^3$ \\
c & $8.11\times 10^{-4}$ & 1.2  &  1.3  &   6.7  &  $6.78\times 10^3$ & $6.87\times 10^3$ & $1.41\times 10^4$ \\
d & $9.43\times 10^{-4}$ & 1.3  &  1.7  &   7.8  &  $9.42\times 10^3$ & $9.64\times 10^3$ & $1.79\times 10^4$ \\
e & $1.07\times 10^{-3}$ & 1.5  &  1.7  &   9.2  &  $1.18\times 10^4$ & $1.17\times 10^4$ & $2.15\times 10^4$ \\
f & $1.16\times 10^{-3}$ & 1.8  &  1.9  &  10.1  &  $1.31\times 10^4$ & $1.30\times 10^4$ & $2.44\times 10^4$ \\
\tableline
\tableline
\end{tabular}
\end{center}
\tablenotetext{a}{see Figure~\ref{fig:3dfull}.}
\end{table*}

\end{document}